\documentclass[prl, twocolumn,amsfonts,amsmath,amssymb,eufrak]{revtex4-1}

\usepackage{epsfig}
\usepackage{color}
\usepackage{bm}
\usepackage{hyperref}
\usepackage[dvipsnames]{xcolor}
\hypersetup{
    bookmarksnumbered=true, 
    unicode=false, 
    pdfstartview={FitH}, 
    pdftitle={}, 
    pdfauthor={}, 
    pdfsubject={}, 
    pdfcreator={}, 
    pdfproducer={}, 
    pdfkeywords={}, 
    pdfnewwindow=true, 
    colorlinks=true, 
    linkcolor=NavyBlue, 
    citecolor=NavyBlue, 
    filecolor=NavyBlue, 
    urlcolor=NavyBlue 
}

\usepackage{braket}

\usepackage{graphicx}
\usepackage{amsthm}

\newcommand{\eq}[1]{\begin{equation} #1 \end{equation}}
\newcommand{\eqa}[2]{\begin{equation} #1 \label{#2} \end{equation}}

\newcommand{\bs}{\boldsymbol}

\newcommand{\figin}[4]
{\begin{figure}[tb]
\centering
\includegraphics[width= #1]{#2.pdf}
\caption{#3}
\label{f:#4}
\end{figure}}

\newcommand{\todayd}{\the\year/\the\month/\the\day}

\newcommand{\lb}{\label}

\newcommand{\Tr}{\mathrm{Tr}}

\newcommand{\bel}{\begin{easylist}}
\newcommand{\eel}{\end{easylist}}

\newcommand{\eref}[1]{Eq.~\eqref{#1}}

\newcommand{\fref}[1]{Fig.~\ref{f:#1}}

\def \({\left(}
\def \){\right)}
\def \[{\left[}
\def \]{\right]}

\newcommand{\abs}[1]{\left|#1\right|}

\newcommand{\sumtwo}[2]%
{\mathop{\sum_{#1}}_{#2}}
\newcommand{\sumthree}[3]%
{\mathop{\mathop{\sum_{#1}}_{#2}}_{#3}}
\newcommand{\sumfour}[4]%
{\mathop{\mathop{\mathop{\sum_{#1}}_{#2}}_{#3}}_{#4}} 
\newcommand{\prodtwo}[2]%
{\mathop{\prod_{#1}}_{#2}}
\newcommand{\mintwo}[2]%
{\mathop{\min_{#1}}_{#2}}
\newcommand{\maxtwo}[2]%
{\mathop{\max_{#1}}_{#2}}
\newcommand{\maxthree}[3]%
{\mathop{\mathop{\max_{#1}}_{#2}}_{#3}}
\newcommand{\limtwo}[2]%
{\mathop{\lim_{#1}}_{#2}}
\newcommand{\suptwo}[2]%
{\mathop{\sup_{#1}}_{#2}}
\newcommand{\supthree}[3]%
{\mathop{\mathop{\sup_{#1}}_{#2}}_{#3}}
\newcommand{\supfour}[4]%
{\mathop{\mathop{\mathop{\sup_{#1}}_{#2}}_{#3}}_{#4}} 
\newcommand{\inftwo}[2]%
{\mathop{\inf_{#1}}_{#2}}
\newcommand{\infthree}[3]%
{\mathop{\mathop{\inf_{#1}}_{#2}}_{#3}}
\newcommand{\inffour}[4]%
{\mathop{\mathop{\mathop{\inf_{#1}}_{#2}}_{#3}}_{#4}} 

\newcommand\calC{{\cal C}}
\newcommand\calD{{\cal D}}
\newcommand\calE{{\cal E}}

\newcommand\calH{{\cal H}}
\newcommand\calI{{\cal I}}

\newcommand\calK{{\cal K}}

\newcommand\calM{{\cal M}}

\newcommand\calS{{\cal S}}

\newcommand{\bsc}{\boldsymbol{c}}

\newcommand{\bsC}{\boldsymbol{C}}

\newcommand{\bszero}{\boldsymbol{0}}

\newcommand{\bbF}{\mathbb{F}}
\newcommand{\bbN}{\mathbb{N}}
\newcommand{\bbO}{\mathbb{O}}
\newcommand{\bbQ}{\mathbb{Q}}
\newcommand{\bbR}{\mathbb{R}}
\newcommand{\bbZ}{\mathbb{Z}}
\newcommand{\ep}{\varepsilon}

\newcommand{\bcs}{\backslash}
\newcommand{\Di}{\mathit{\Delta}}

\newcommand{\para}[1]{{\em #1}\/.---}

\newtheorem{thm}{Theorem}
\newtheorem{lm}[thm]{Lemma}
\newtheorem{pro}[thm]{Proposition}
\newtheorem{cor}[thm]{Corollary}

\newcommand{\bthm}[1]{\begin{thm} #1 \end{thm}}
\newcommand{\blm}[1]{\begin{lm} #1 \end{lm}}
\newcommand{\bpro}[1]{\begin{pro} #1 \end{pro}}

\theoremstyle{definition}
\newtheorem{dfn}[thm]{Definition}
\newcommand{\bdf}[1]{\begin{dfn} #1 \end{dfn}}
\newtheorem{cjt}[thm]{Conjecture}

\newcommand{\bsDi}{\bs{\Di}}
\newcommand{\frR}{\mathfrak{R}}

\def\rnum#1{\resizebox{0.5em}{\height}{\expandafter{\romannumeral #1}}}
\def\Rnum#1{\resizebox{0.5em}{\height}{\uppercase\expandafter{\romannumeral #1}}}

\newcommand{\sbar}{\;\rule{0pt}{9.5pt}\right|\;}
\newcommand{\lset}{\left\{\left.}
\newcommand{\rset}{\right\}}
\newcommand{\ketbra}[2]{|{#1}\rangle\!\langle{#2}|}
\newcommand{\dm}[1]{\ketbra{#1}{#1}}
\newcommand{\bal}{\begin{equation}\begin{aligned}}
\newcommand{\eal}{\end{aligned}\end{equation}}
\usepackage{titlesec}
\usepackage[lmargin=.7in,rmargin=.7in,tmargin=.7in,bmargin=1in]{geometry}

\begin{document}


\title{
Arbitrary Amplification of Quantum Coherence in Asymptotic and Catalytic Transformation}

\author{Naoto Shiraishi} 
\email{shiraishi@phys.c.u-tokyo.ac.jp}
\affiliation{Department of Basic Science, The University of Tokyo, 3-8-1 Komaba, Meguro-ku, Tokyo 153-8902, Japan}%

\author{Ryuji Takagi} 
\email{ryujitakagi.pat@gmail.com}
\affiliation{Department of Basic Science, The University of Tokyo, 3-8-1 Komaba, Meguro-ku, Tokyo 153-8902, Japan}%

\begin{abstract}
Quantum coherence is one of the fundamental aspects distinguishing classical and quantum theories. Coherence between different energy eigenstates is particularly important, as it serves as a valuable resource under the law of energy conservation. A fundamental question in this setting is how well one can prepare good coherent states from low coherent states and whether a given coherent state is convertible to another one. Here, we show that any low coherent state is convertible to any high coherent state arbitrarily well in two operational settings: asymptotic and catalytic transformations. For a variant of asymptotic coherence manipulation where one aims to prepare desired states in local subsystems, the rate of transformation becomes unbounded regardless of how weak the initial coherence is. In a non-asymptotic transformation with a catalyst, a helper state that locally remains in the original form after the transformation, we show that an arbitrary state can be obtained from any low coherent state. Applying this to the standard asymptotic setting, we find that a catalyst can increase the coherence distillation rate significantly---from zero to infinite rate. We also prove that such anomalous transformation requires small but non-zero coherence in relevant modes, establishing the condition under which a sharp transition of the operational capability occurs. Our results provide a general characterization of the coherence transformability in these operational settings and showcase their peculiar properties compared to other common resource theories such as entanglement and quantum thermodynamics.
\end{abstract}

\maketitle

Quantum coherence between different energy eigenstates is a valuable resource inevitable for quantum clocks~\cite{JB03}, metrology~\cite{GLM06}, and work extraction~\cite{LJR15}.
Under the law of energy conservation, coherence in the above sense is easily lost due to decoherence, while it is impossible to create and inflate coherence without any help.
In this regard, coherence is a precious quantum resource that should be utilized as efficiently as possible.

A central problem concerning quantum coherence as an operational resource is to characterize its manipulability with energy-conserving unitary~\cite{GS08,Mar-thesis}. 
This physical setting comes with a fundamental constraint that the total amount of quantum coherence cannot be increased by energy-conserving operations. 
To understand its manipulation power, two formalisms of state transformations---asymptotic and catalytic transformations---have been actively investigated. 

One standard setting for resource manipulation is the \emph{asymptotic transformation}, where one aims to convert many copies of the initial quantum state to many copies of another target state~\cite{Wilbook}.
The key performance quantifier of the asymptotic manipulation is its transformation rate, the ratio of the number of copies of the final state to those of the initial one.  
On the asymptotic coherence manipulation, it has been shown that there is a strong limitation that the transformation rate from generic mixed states to pure coherent states is zero~\cite{Mar20}.

Another standard setting for resource manipulation is \emph{catalytic transformation}, where one is allowed to borrow the help of another auxiliary system called catalyst---an ancillary system that should return to its own state at the end of the process.
In particular, \emph{correlated catalyst}, which could have a correlation with the main system after the transformation, has shown to be effective in enhancing the resource manipulability for several physical settings~~\cite{Wilming2017axiomatic,  Muller2018correlating, Boes2019vonNeumann, Shiraishi2021quantum,  Lie2021catalytic, Wilming2021entropy, Lipka-Bartosik2021catalytic, Kondra2021catalytic, Rubboli2022fundamental, Wilming2022correlations, Yadin2022catalytic,  Lami2023catalysis, Ganardi2023catalytic}. 
However, similarly to the asymptotic transformation, fundamental limitations on catalytic enhancement have been observed.
A notable result is the coherence no-broadcasting theorem~\cite{Lostaglio2019coherence_asymmetry,Marvian2019no-broadcasting}, showing that no coherence could be created with a correlated catalyst if the input state is exactly incoherent.    
These previous studies, both on asymptotic and catalytic transformations, indicate the potential difficulty of manipulating quantum coherence.

Contrarily to these suggestions, we here show that an arbitrary coherence manipulation is enabled in asymptotic and catalytic coherence transformation. 
We consider a variant of the asymptotic transformation where one aims to prepare a target state on each subsystem~\cite{Fer23} and show that the transformation rate becomes unbounded if the initial state has non-zero coherence. 
In the correlated-catalytic transformation, we prove that arbitrary state transformation becomes possible as long as the initial state has non-zero coherence. 
This shows that the observation from the coherence no-broadcasting theorem is unstable about the perturbation of the initial state in the following sense: As long as the initial state contains even a tiny amount of coherence, every coherent state suddenly becomes reachable.
In addition, for target states besides measure-zero exceptions, \emph{exact} transformation is possible, which is a much stronger claim than the conventional resource-theoretic results allowing a small error in the final state.

As a direct consequence of our result, we show that the standard asymptotic transformation rate becomes infinite with the help of correlated catalysts.
This resolves the open problem proposed in Ref.~\cite{Lami2023catalysis}, asking whether correlated catalysts could improve the asymptotic rate at all, in the most drastic manner---catalysts can make undistillable coherent states infinitely distillable.

Our protocols require a non-zero amount of coherence---even if extremely small---in the initial state to implement arbitrary state conversions.
To fully characterize this requirement, we formalize no-go theorems on state conversions by introducing the notion of \emph{resonant coherent modes}.
These no-go theorems reveal that initial coherence, even if it is negligibly small, is inevitable for arbitrary state conversions, and exactly zero coherence must result in zero coherence.
Together with the feasible transformations described above, these characterize state transformability both in asymptotic and catalytic settings, revealing that the distinction between zero and non-zero coherence is an extremely sharp threshold.

We remark that these ``amplification'' effects do not contradict the physical requirements that the total amount of coherence should not increase. Our results rest on the fact that coherence can \emph{locally} increase, as observed in several settings previously~\cite{Manzano2019autonomous,Ding2021amplifying_asymmetry}.  
Our results extend these observations in the context of asymptotic and catalytic coherence manipulation and provide general characterizations of the anomalous coherence amplification phenomena observed in each operational setting.

\para{Coherence transformation}
Superposition between energy eigenstates is manifested in time evolution. 
For a system with Hamiltonian $H$, a state $\rho$ is called \emph{coherent} if $\mathcal{U}_t(\rho)\neq \rho$ for some time $t$, where $\mathcal{U}_t(\rho):= e^{-iHt}\rho e^{iHt}$ is the unitary time evolution. A state is called \emph{incoherent} if it is not coherent. 
We remark that the coherence we consider in this work is what is so-called \emph{unspeakable coherence} \cite{Marvian2016howto}. (Not to be confused with another type known as \emph{speakable coherence}~\cite{Baumgratz2014quantifying}.)

Available operations in manipulating quantum coherence should not create coherence from incoherent states, as respecting the law of energy conservation. 
In reflecting this restriction, a natural set of available operations for the coherence manipulation is the \emph{covariant operations} with time translation~\footnote{In the context of resource theories, covariant operations correspond to the class of completely resource non-generating operations (see also Proposition~\ref{pro:covariant_app} in the Supplemental Material), which constitutes the standard set of free operations considered for the resource theory of unspeakable coherence.}, i.e., the action of a channel $\Lambda:S\to S'$, where $S$ and $S'$ are input and output systems, commutes with the unitary time evolution as $\Lambda\circ \mathcal{U}_t^S =\mathcal{U}_t^{S'}\circ \Lambda$ for all $t$~\cite{CG19,GS08,Mar-thesis}.
From the operational perspective, any covariant operation $\Lambda$ can be equivalently written by an energy-conserving unitary $U$ and an incoherent state $\eta$ as $\Lambda(\rho)=\Tr_{A}[U(\rho\otimes\eta)U^\dagger]$, where $A$ is some auxiliary system~\cite{Keyl1999optimal,Mar-thesis}. 
In other words, covariant operations are operations which can be implemented by an energy-conserving unitary with incoherent states.

\para{Coherent modes}
Our findings clarify that whether relevant modes have (maybe tiny but) nonzero coherence leads to a drastic change.
To formalize this, we introduce the notion of resonant coherent modes.
A mode for $\Di$ is a pair of two energy levels with energy difference $\Di$, and a state $\rho$ has a coherent mode $\Delta$ if $\rho_{ij}\neq 0$ with $E_i-E_j=\Delta$ is satisfied for some $i,j$, where $\rho_{ij}:=\braket{i|\rho|j}$ and $\ket{i}$ is an energy eigenstate with energy $E_i$ for the given Hamiltonian $H$.
We then define the set $\calC(\rho)$ of resonant coherent modes of state $\rho$ as all linear combinations of non-zero coherent modes with integer coefficients, i.e., 
\eq{
\calC(\rho):=\lset x \sbar x=\sum_{{i,j \ (\rho_{ij}\neq 0)}} n_{ij}\Di _{ij},\, n_{ij}\in \bbZ \rset
}
for an energy interval $\Di_{ij}=E_i-E_j$ and a non-zero off-diagonal entry $\rho_{ij}$ of a density matrix $\rho$ for energies $E_i$ and $E_j$.
Notably, in the asymptotic and catalytic coherence manipulation, one can create a coherence on mode $\Di=\Di_1+\Di_2$ if the initial state has coherence on modes $\Di_1$ and $\Di_2$~\cite{TS22}.

\para{Asymptotic manipulation}
We first consider the asymptotic manipulation. 
Suppose $\rho$ is an initial state and $\rho'$ is a target state. 
In the standard framework of asymptotic transformation, one considers a series $\{\Lambda_n\}_n$ of available operations that transforms $\rho^{\otimes n}$ to $\rho'^{\otimes \lfloor rn \rfloor}$ with vanishing error at the limit of $n\to\infty$. 
The asymptotic transformation rate
$R(\rho\to\rho')$ is the supremum over all achievable rates $r$. 
When $\rho'$ is a pure state $\phi$, it is particularly called asymptotic distillation. For coherence distillation with covariant operations, the distillation rate $R(\rho\to\phi)$ is known to be zero for an arbitrary full-rank state $\rho$ and an arbitrary coherent pure state $\phi$~\cite{Mar20}, which puts a fundamental restriction on the tangibility of coherence as an operational resource.

\figin{8.5cm}{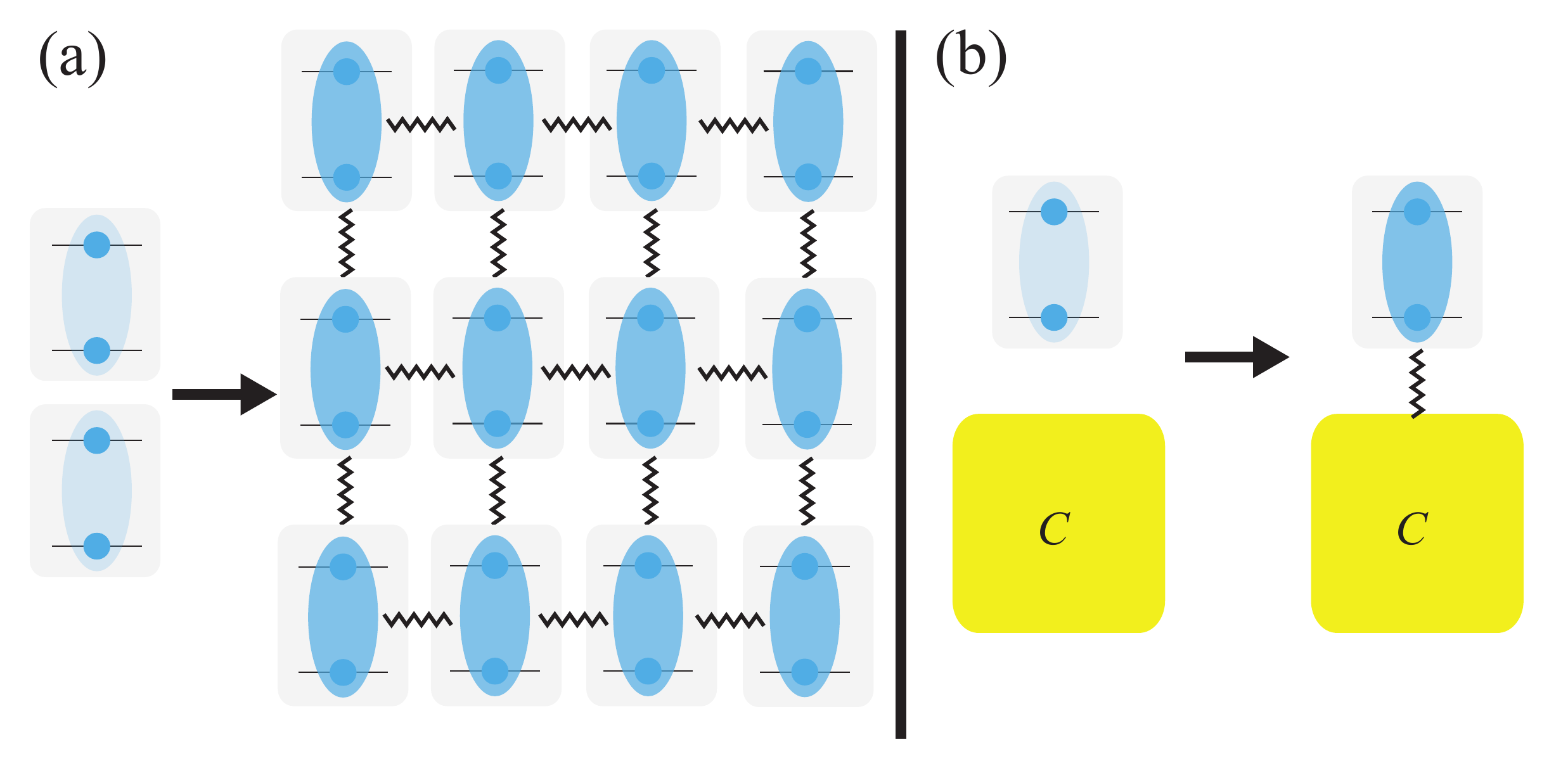}{
(a): An asymptotic marginal transformation maps $n$ copies of $\rho$ into $m$ copies of $\rho'$ with correlation among copies.
Theorem~\ref{thm:asymptotic} states that for almost all $\rho$ and $\rho'$, the rate of asymptotic marginal transformation of $\rho\to \rho'$ defined as $\lim_{n\to \infty} \max_m \frac mn$ is unbounded.
(b): A correlated-catalytic transformation maps a product state of system $S$ and catalyst $C$ written as $\rho\otimes c$ to $\tau$ such that $\Tr_S[\tau]=c$ and $\Tr_C[\tau]=\rho'$.
Theorem~\ref{thm:correlated catalyst achievability} states that for almost all $\rho$ and $\rho'$, $\rho$ is convertible to $\rho'$ with a correlated catalyst.
In addition, the strength of the correlation between the main and catalytic systems can be made arbitrarily small.
}{asymptotic-catalytic}

However, the necessity of obtaining the state close to $\phi^{\otimes \lfloor rn \rfloor}$ can be reasonably relaxed for many operational settings. 
For instance, consider the scenario where multiple parties are separated from each other and would like to consume a good coherent state locally. In such a setting, the quality of the resource state is determined by how close the local marginal state is to the maximally coherent state.
The framework that fits this operational setting was considered previously and called \emph{asymptotic marginal transformation}~\cite{Fer23,Ganardi2023catalytic}. 
Suppose $\rho$ and $\rho'$ are the states on the systems $S$ and $S'$ respectively. The state $\rho$ can be converted to $\rho'$ with an asymptotic marginal transformation with rate $r$ if there exists a series of available operations $\{\Lambda_n\}_n$ from $S^{\otimes n}$ to $S'^{\otimes \lfloor r n\rfloor}$ such that the reduced state of $\Lambda_n\left(\rho^{\otimes n}\right)$ on every subsystem approaches $\rho'$ with a vanishing error at the $n\to\infty$ limit. 
If the reduced state of any single subsystem exactly coincides with the target state $\rho'$ for some finite $n$, we say that this asymptotic marginal transformation is \emph{exact}.
Although the asymptotic marginal transformation rate $\tilde R(\rho\to\rho')$, which is defined as the highest achievable rate in the marginal asymptotic conversion, serves as an upper bound of the standard asymptotic transformation $\tilde R(\rho\to\rho')\geq R(\rho\to\rho')$, these two rates coincide in many settings such as entanglement, quantum thermodynamics, and non-classicality~\cite{Fer23}, suggesting that this relaxation may not realize a significant improvement in the ability of transformation (see also Sec.~\ref{subsec:asymptotic preliminaries} in the Supplemental Material~\footnote{See the Supplemental Material for detailed proofs and discussions of our main results, which includes Refs.~\cite{Bennett1996concentrating,Bravyi2005universal,Brandao2013resource,Horodecki2013fundamental,Horodecki2009quantum,Yadin2018operational,Kwon2019nonclassicality,Genoni2010quantifying,Takagi2018convex,Albarelli2018resource,Veitch2014resource,Howard2017application,Takagi2019operational,Takagi2019general,Uola2019quantifying,Regula2021operational,Anshu2018quantifying,Regula2018convex,Liu2019resource,Gour2019how,Gour2020dynamical_resources,Gonda2023monotones_general,Horodecki2013quantumness,Gour2017quantum,Liu2019one-shot,Regula2020benchmarking,Kuroiwa2020general,Fang2020nogo-state,Liu2020operational,Fang2020no-go,Regula2020oneshot,Regula2021fundamental,Regula2022functional,Regula2022tight_constraints,Takagi2022one-shot_yield-cost,Regula2022probabilistic,Audenaert2003entanglement,Regula2019one-shot,Wang2020cost,Cirac1999optimal,Marvian2022operational,Zhang2017detecting,Marvian2014extending,Tak19,Hansen2008metric,Synak-Radtke2006on_asymptotic,Coladangelo2019additive,Jonathan1999entanglement-assisted,Bu2016catalytic,Campbell2011catalysis,Marvian2013theory_manipulations,Hickey2018quantifying}}). 
\nocite{Bennett1996concentrating,Bravyi2005universal,Brandao2013resource,Horodecki2013fundamental,Horodecki2009quantum,Yadin2018operational,Kwon2019nonclassicality,Genoni2010quantifying,Takagi2018convex,Albarelli2018resource,Veitch2014resource,Howard2017application,Takagi2019operational,Takagi2019general,Uola2019quantifying,Regula2021operational,Anshu2018quantifying,Regula2018convex,Liu2019resource,Gour2019how,Gour2020dynamical_resources,Gonda2023monotones_general,Horodecki2013quantumness,Gour2017quantum,Liu2019one-shot,Regula2020benchmarking,Kuroiwa2020general,Fang2020nogo-state,Liu2020operational,Fang2020no-go,Regula2020oneshot,Regula2021fundamental,Regula2022functional,Regula2022tight_constraints,Takagi2022one-shot_yield-cost,Regula2022probabilistic,Audenaert2003entanglement,Regula2019one-shot,Wang2020cost,Cirac1999optimal,Marvian2022operational,Zhang2017detecting,Marvian2014extending,Tak19,Hansen2008metric,Synak-Radtke2006on_asymptotic,Coladangelo2019additive,Jonathan1999entanglement-assisted,Bu2016catalytic,Campbell2011catalysis,Marvian2013theory_manipulations,Hickey2018quantifying}

Despite these previous observations, we prove that any low coherent state can be transformed to any high coherent state with an arbitrarily high transformation rate.
Namely, there is no restriction on coherence transformation, and all states without measure-zero exceptions admit infinite asymptotic marginal distillation rates (\fref{asymptotic-catalytic}.(a)). 

\begin{thm}\label{thm:asymptotic}
For arbitrary states $\rho$ and $\rho'$, $\tilde R(\rho\to\rho')$ diverges if $\rho$ has non-zero coherence in the sense of $\mathcal{C}(\rho')\subseteq\mathcal{C}(\rho)$. Moreover, the asymptotic marginal transformation can be made exact if $\rho'$ is full rank. In both cases, the correlation between one subsystem and the others can be made arbitrarily small.
On the other hand, if $\mathcal{C}(\rho')\not\subseteq\mathcal{C}(\rho)$, even a single copy of $\rho'$ cannot be prepared from any number of copies of $\rho$ with arbitrarily small error. 
\end{thm}

We remark that $\mathcal{C}(\rho')\subseteq\mathcal{C}(\rho)$ is quite a mild condition since any state $\rho$ with extremely small but non-zero coherence on all modes automatically passes this requirement regardless of $\rho'$.

Theorem~\ref{thm:asymptotic} provides the complete characterization of the general asymptotic marginal coherence transformation, including the case of distillation when $\rho'$ is pure.
Intuitively speaking, if the initial state contains non-zero coherence on modes that are coherent in the target state, then an arbitrary rate can be realized. 
The condition $\calC(\rho')\subseteq \calC(\rho)$, whether the state has (maybe extremely small but) non-zero coherence or has exactly zero coherence, serves as an extremely sharp and the only threshold separating infinite and zero asymptotic transformation rates.

The diverging rate for the exact transformation shown in Theorem~\ref{thm:asymptotic} is also remarkable.
In fact, asymptotic resource transformation typically comes with a severe restriction when no errors are allowed, and therefore much less is known for exact transformation compared to transformation with a vanishing error.
Our result presents a rare scenario in which exact transformation realizes an outstanding performance that coincides with the performance of non-zero error transformation.

\para{Correlated-catalytic transformation}
We now consider the correlated-catalytic transformation, where we employ an auxiliary system $C$ called catalyst which does not change its own state between the initial and the final state but helps state conversion in system $S$.
We say that $\rho$ is convertible to $\rho'$ through correlated-catalytic transformation if there exists a finite-dimensional catalytic system $C$ with a catalyst state $c$, and a covariant operation $\Lambda$ on $SC$ such that $\tau=\Lambda(\rho\otimes c)$ with $\Tr_S[\tau]=c$ and $\Tr_C[\tau]=\rho'$.
Our final state may have a correlation between the system and the catalyst, which reflects the name ``correlated catalyst".

We investigate covariant operations with a correlated catalyst.
Recent studies have revealed a severe limitation for correlated-catalytic covariant operations, called the coherence no-broadcasting theorem~\cite{Marvian2019no-broadcasting, Lostaglio2019coherence_asymmetry}.
This theorem states that a fully incoherent initial state is convertible only to an incoherent state through a covariant operation even with the help of a correlated catalyst.
This may suggest that a correlated catalyst offers little advantage in state convertibility with quantum coherence.
However, we show exactly the opposite---correlated catalysts allow enormous operational power to most covariant state conversions, and the only exception is the case with no coherence in the initial state.

\begin{thm}\label{thm:correlated catalyst achievability}
For arbitrary states $\rho$ and $\rho'$, $\rho$ is convertible to $\rho'$ with a correlated catalyst with an arbitrarily small error if $\mathcal{C}(\rho')\subseteq\mathcal{C}(\rho)$, and the transformation can be made exact if $\rho'$ is full rank.
In addition, the correlation between the system and catalyst can be made arbitrarily small. 
\end{thm}

This shows that a correlated catalyst enables an arbitrary coherence amplification---an almost incoherent state can be transformed to an almost maximally coherent state with a correlated catalyst (\fref{asymptotic-catalytic}.(b)), solving the conjecture in Ref.~\cite{TS22} in the affirmative.
Similarly to the case of asymptotic transformation, the only meaningful distinction lies in whether the state has non-zero coherent modes or not.
Notably, the correlation between the system and the catalyst can be made arbitrarily small, implying that the final state $\tau$ is extremely close to a product state of $\rho'\otimes c$.

By choosing the initial state as $\rho^{\otimes n}$ and the target state as $\phi^{\otimes rn}$ for a coherent state $\rho$ and a pure coherent state $\phi$, there exists a correlated catalyst that enables the transformation from $\rho^{\otimes n}$ to $\phi^{\otimes rn}$ with an arbitrarily small error for every $n$ and $r$. 
This setup corresponds to the standard (not marginal) asymptotic distillation, i.e., the error in the final state is measured for the entire state $\phi^{\otimes rn}$, assisted by correlated catalysts.
Noting that the standard asymptotic distillation rate $R(\rho\to\phi)$ without a catalyst is zero for every full-rank state $\rho$~\cite{Mar20}, our result gives the first example for which the catalyst improves the asymptotic transformation rate, resolving the open problem raised in Ref.~\cite{Lami2023catalysis}.

As for the converse, we expect that $\calC(\rho')\subseteq \calC(\rho)$ also gives the necessary condition for the transformation to exist. 
Here, we give a partial result toward the full solution to this problem. 
As naturally guessed, coherence in the initial state would not be helpful in creating coherence on the mode that is only irrationally related to the resonant coherent modes.
To formalize this, we introduce $\calC'$, which is an extension of $\calC$ to rational coefficients: $\calC'(\rho):=\{ x| x=\sum_{{i,j \ (\rho_{ij}\neq 0)}} n_{ij}\Di _{ij}, n_{ij}\in \bbQ\}$.
We then obtain the necessary condition for the approximate correlated-catalytic covariant transformation (\fref{no-go-2}).

\begin{thm}\label{thm:correlated catalyst converse}
For two states $\rho$ and $\rho'$ such that $\calC'(\rho')\nsubseteq \calC'(\rho)$, there does not exist a correlated-catalytic covariant transformation from $\rho$ to $\rho'$.
\end{thm}

\figin{8.5cm}{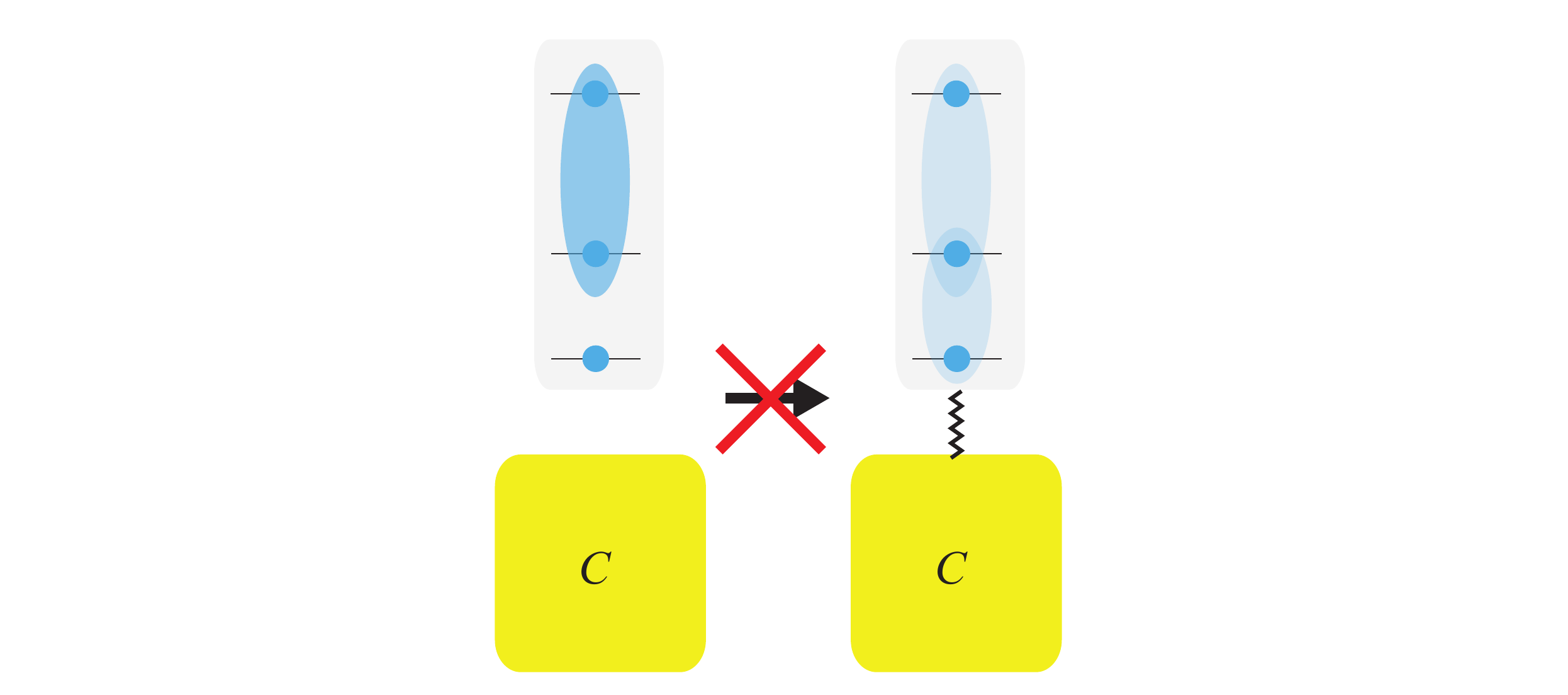}{
Suppose that a mode has no coherence in the initial state.
Then even if the initial state has coherence on other modes irrationally related to the mode in interest, a covariant operation with a correlated catalyst cannot provide coherence on this mode.
This restriction is stronger than the coherence no-broadcasting theorem~\cite{Marvian2019no-broadcasting, Lostaglio2019coherence_asymmetry}.
}{no-go-2}

This result can be understood as \emph{mode no-broadcasting}---new coherent modes cannot be created by a covariant operation with a correlated catalyst.
This contains the coherence no-broadcasting theorem as a special case with $\mathcal{C}'(\rho)=\{0\}$ and extends it to the case of coherent initial states.
We conjecture that the above condition is strengthened to $\mathcal{C}(\rho')\not\subseteq\mathcal{C}(\rho)$, which would provide the exact characterization of the feasible coherence transformation with correlated catalyst together with Theorem~\ref{thm:correlated catalyst achievability}.

\para{Proof sketch}
Here, we provide a proof sketch of our main results.
The complete proofs are presented in the Supplemental Material.

We first outline the proof of the achievable part of Theorem~\ref{thm:asymptotic}. 
Our protocol employs another operational framework known as marginal-catalytic transformations~\cite{TS22} (see also Ref.~\cite{Ding2021amplifying_asymmetry}). 
In particular, it was shown that for any full-rank state $\rho'$ there exists a set $C_1,\ldots , C_N$ of catalytic systems with states $c_1,\ldots , c_N$ and a covariant operation $\Lambda$: $S\otimes C_1\otimes \cdots \otimes C_N\to S'\otimes C_1\otimes \cdots \otimes C_N$ such that $\Lambda(c_1\otimes \cdots \otimes c_N)=\tau$ with $\Tr_{C_1,\ldots , C_N}[\tau]=\rho'$ and $\Tr_{\backslash C_i}[\tau]=c_i$ for all $i=1,\ldots , N$.
Furthermore, these catalysts are partially reusable:
If we have $N^k$ sets of catalysts $c_1\otimes\cdots\otimes c_N$, an appropriate recombination of them allows one to prepare $(k+1)N^k$ copies of $\rho'$ with these marginal catalysts.

We now construct our protocol, which is inspired by Ref.~\cite{Ganardi2023catalytic}.
We first show that a set $c_1\otimes\cdots \otimes c_N$ of catalysts can be prepared exactly from $\rho^{\otimes \mu}$ by a covariant operation for some integer $\mu$. 
This allows us to transform $\mu N^k$ copies of $\rho$ into $N^k$ sets of catalysts.
Using these catalysts, we obtain $(k+1)N^k$ copies of $\rho'$ by a marginal catalytic covariant transformation, after which we discard the catalytic systems. 
The transformation rate is $(k+1)/\mu$, which can be made arbitrarily large by setting sufficiently large $k$.
This transformation can be made exact for a full-rank target state $\rho'$ by employing the result in Ref.~\cite{Wilming2022correlations}.

The converse part of Theorem~\ref{thm:asymptotic} can be shown by utilizing the properties of the modes of asymmetry~\cite{Marvian2014modes}.

Theorem~\ref{thm:correlated catalyst achievability} can be obtained by applying the well-known technique to derive correlated-catalytic convertibility from asymptotic convertibility with vanishing error.
This type of result was first shown in Ref.~\cite{Shiraishi2021quantum} in the context of quantum thermodynamics (c.f. Refs.~\cite{Dua05,AN} for exact asymptotic transformation), and a general form of statement is explicitly shown and proven in Ref.~\cite{TS22}.
In particular, this construction was recently used to convert asymptotic marginal transformation to correlated-catalytic transformation~\cite{Wilming2022correlations}.

We finally outline the proof of Theorem~\ref{thm:correlated catalyst converse}.
We suppose contrarily that the final state $\rho'$ has coherence on a mode $\Delta E\not\subseteq\calC'(\rho)$ and derive contradiction with the coherence no-broadcasting theorem~\cite{Marvian2019no-broadcasting, Lostaglio2019coherence_asymmetry}.
Let $L(\Di)$ be an infinite-dimensional system whose energy levels form a ladder with energy interval $\Di$.
We embed the main system $S$ and catalytic system $C$ into a product of ladder systems $L(\Di_0)\otimes L(\Di_1)\otimes L(\Di_2)\otimes \cdots$ such that $\Di_0$ is an integer multiple of $\Delta E$ ($\Delta E = m\Di_0$ for some integer $m$), and the set $\{\Di_0,\Di_1,\Di_2,\dots\}$ are rational-linearly independent.
By assumption, $\rho'$ has coherence in $L(\Di_0)$, for which $\rho$ is incoherent.
For brevity, we abbreviate the set $\Di_1,\Di_2,\ldots $ as $\tilde\bsDi$.

Our key observation is that since a covariant operation acts on each rational-linearly independent mode separately, if $\rho\otimes c$ on $L(\Di_0)\otimes L(\tilde\bsDi)$ can be transformed to $\tau$ by a covariant operation, the same transformation is also possible on systems with another arbitrary set $\tilde\bsDi'=(\Di'_1, \Di'_2,\ldots )$.
Namely, $\rho\otimes c\to \tau$ on $L(\Di_0)\otimes L(\tilde\bsDi')$ (the same density matrix on ladders with different energy spacings) is possible by a covariant operation.
Setting $\tilde\bsDi'=\bszero$ in the above modification, where all states outside $L(\Di_0)$ are degenerate, we find that $\rho$ on $L(\Di_0)\otimes L(\bszero)$ is completely incoherent.
On the other hand, the final state $\rho'$ has coherence in $L(\Di_0)$, which contradicts the coherence no-broadcasting theorem.

\para{Discussion}
We showed the anomalous potential of the manipulation of quantum coherence in the asymptotic and catalytic coherence distillation.
These results are highly special to quantum coherence that cannot be seen in other resource theories such as entanglement~\cite{CW04, AF04}, quantum thermodynamics~\cite{Wilming2017axiomatic}, and speakable coherence~\cite{Baumgratz2014quantifying, XLF15, Marvian2016howto} (see Sec.~\ref{sec:general resource theories} in the Supplemental Material).
Related to this, we stress that our result is different from the well-known embezzlement phenomena observed in several resource theories~\cite{vanDam2003universal_entanglement,Brandao15second_laws}, admitting arbitrary state conversions by allowing a small error in a catalyst. 
Our framework allows no errors in the catalyst, and thus the operational capability comes from an entirely different mechanism. 

Our results shed light on the power of correlation in resource manipulation.
In fact, without correlation, amplification of coherence is impossible in both asymptotic and catalytic settings. 
The importance of correlation has already been discussed intensively in the context of quantum thermodynamics~\cite{Lostaglio2015stochastic, MP, SCR, LN23}.
Quantum thermodynamics with an uncorrelated catalyst has many restrictions with R\'{e}nyi entropies in state convertibility~\cite{AN, Klimesh2007inequalities, Turgut2007catalytic_transformations, Brandao15second_laws}, while most of the restrictions are lifted by proper use of correlations and only the second law of thermodynamics with the relative entropy remains~\cite{Muller2018correlating, Shiraishi2021quantum}.
For the coherence transformation, previous studies~\cite{Ding2021amplifying_asymmetry, TS22} showed an astonishing operational power enabled by correlations between multiple catalysts. 
Our results confirm that the unbounded power of coherence transformation is also present in the setting with much more operational motivation---asymptotic and correlated-catalytic coherence transformation---lifting quantum coherence as an even more tangible operational resource.

\let\oldaddcontentsline\addcontentsline
\renewcommand{\addcontentsline}[3]{}

\begin{acknowledgments}
{\it Acknowledgments.} ---
We thank Eunwoo Lee for discussions on group representations, and Kohdai Kuroiwa for the asymptotic continuity. 
N.S. was supported by JSPS Grants-in-Aid for Scientific Research Grant Number JP19K14615. R.T. is supported by JSPS KAKENHI Grant Number JP23K19028.

\end{acknowledgments}

\let\addcontentsline\oldaddcontentsline

{\it Note added.} --- During the completion of our manuscript, we became aware of an independent related work by Kondra et al.~\cite{Kondra2023correlated}, which was concurrently posted to arXiv with ours.
Also, an anonymous referee of the QIP conference notified us that when $\rho'$ is pure and the period (the minimum time after which the state returns to the original one) for $\rho$ and $\rho'$ coincide, one can also obtain the diverging asymptotic marginal transformation rate (with an arbitrary small error) by generalizing the construction for sublinear coherence distillation in Ref.~\cite[Supplementary Note 7]{Mar20} to the case of marginal asymptotic conversion. 
This approach, which is different from ours, in fact admits a larger target state, up to the size sublinear in the number of copies of $\rho$.
Our Theorem~\ref{thm:asymptotic}, on the other hand, applies to the fully general setting and contains further insights into the possibility of exact transformation and fundamental limitations imposed by the resonant coherence modes. 
We thank the referee for their insightful comments.


\let\oldaddcontentsline\addcontentsline
\renewcommand{\addcontentsline}[3]{}

\bibliographystyle{apsrmp4-2}
\bibliography{myref}

\let\addcontentsline\oldaddcontentsline

\onecolumngrid


\newcommand{\vo}{\upsilon}
\newcommand{\midskip}{\vspace{3pt}}
\setcounter{thm}{0}
\renewcommand{\thethm}{S.\arabic{thm}}
\renewcommand{\thedfn}{S.\arabic{dfn}}
\renewcommand{\thelm}{S.\arabic{lm}}
\renewcommand{\thepro}{S.\arabic{pro}}

\titleformat{\section}
  {\centering\normalfont\bfseries}
  {Appendix \thesection:}{.5em}{}
\titleformat{\subsection}
  {\centering\normalfont\bfseries}
  {\thesection.\thesubsection}{.5em}{}
\setcounter{equation}{0}
\setcounter{secnumdepth}{5}

\makeatletter
\def\shorttableofcontents#1#2{\bgroup\c@tocdepth=#2\@restonecolfalse
  \settowidth\js@tocl@width{\headfont\prechaptername\postchaptername}%
  \settowidth\@tempdima{\headfont\appendixname}%
  \ifdim\js@tocl@width<\@tempdima \setlength\js@tocl@width{\@tempdima}\fi
  \ifdim\js@tocl@width<5zw \divide\js@tocl@width by 5 \advance\js@tocl@width 4zw\fi
\if@tightshtoc
    \parsep\z@
  \fi
  \if@twocolumn\@restonecoltrue\onecolumn\fi
  \@ifundefined{chapter}%
  {\section*{{#1}
        \@mkboth{\uppercase{#1}}{\uppercase{#1}}}}%
  {\chapter*{{#1}
        \@mkboth{\uppercase{#1}}{\uppercase{#1}}}}%
  \@startshorttoc{toc}\if@restonecol\twocolumn\fi\egroup}
\makeatother

\newpage
\newgeometry{hmargin=1.2in,vmargin=0.8in}

\setcounter{page}{1}

\begin{center}
{\large \bf Supplemental Material\\  \vspace{0.2cm} ``Arbitrary Amplification of Quantum Coherence in Asymptotic and Catalytic Transformation"}\\
\vspace*{0.5cm}
Naoto Shiraishi and Ryuji Takagi  \\
\vspace*{0.1cm}

{\it Department of Basic Science, The University of Tokyo} 

\end{center}

\tableofcontents

\section{General setups}

\subsection{Preliminaries}
\subsubsection{Quantum resource theories}

The coherence transformation discussed in this work can be understood in a broader context of quantum resource theories~\cite{CG19}, which is a general framework that describes the quantification and feasible manipulation of quantities that are considered ``precious'' under given physical settings.
The physical restriction can be formalized by specifying a set $\bbF$ of \emph{free states} and a set $\bbO$ of \emph{free operations} that are assumed to be freely accessible in the setting of interest.
The requirement imposed on free operations is that they should not be able to create resourceful (i.e., non-free) states from free states, i.e., every free operation $\Lambda\in\bbO$ satisfies $\Lambda(\sigma)\in\bbF$ for all $\sigma\in\bbF$, justifying the notion of ``free'' operations. 

These concepts not only allow us to characterize the resourceful states but also motivate us to \emph{quantify} the amount of resourcefulness contained in a given resourceful state $\rho\not\in\bbF$. 
The resource quantification can be formalized by \emph{resource measures}, also known as \emph{resource monotones}, which are functions from quantum states to real numbers that return the ``amount'' of resourcefulness.
For a function $R$ to be a valid resource measure, it is required that (1) it takes the minimum value for free states: $R(\sigma)=c$ for all $\sigma\in\bbF$ with some constant $c$ and $R(\rho)\geq c$ for every state $\rho$, (2) it does not increase under free operations (monotonicity): $R(\rho)\geq R(\Lambda(\rho))$ is satisfied for all free operations $\Lambda\in\bbO$ and all states $\rho$.

The fundamental problem in any operational setting is to characterize the feasible state transformation by a quantum process accessible in a given setting.
The resource theory framework allows us to formulate this question as follows: for given states $\rho$ and $\rho'$, does there exist a free operation $\Lambda\in\bbO$ such that $\Lambda(\rho)=\rho'$? 
We call this type of state transformation enabled by a free operation \emph{free transformation}. 
The problem of feasible state transformation is closely related to the notion of resource quantification since the monotonicity of a resource measure $R$ ensures $R(\rho)\geq R(\Lambda(\rho))$. 
Therefore, $R(\rho)\geq R(\rho')$ serves as a necessary condition for the free transformation from $\rho$ to $\rho'$ to be possible. 
This also reflects the intuition that ``free operations should not increase the amount of precious resources''. 
On the other hand, proving the sufficient part requires one to construct a specific free operation $\Lambda$ that actually transforms $\rho$ to $\rho'$. 
The ultimate goal then is to obtain the necessary and sufficient conditions for feasible free transformation for arbitrary given states $\rho$ and $\rho'$.

The performance of free transformation can be studied in various settings, each of which comes with an operational motivation. 
In this work, we mainly focus on two standard settings known as \emph{asymptotic} and \emph{catalytic} transformation, which we discuss in Sec.~\ref{sec:asymptotic} and Sec.~\ref{sec:catalytic} respectively.

The characterization of feasible free transformation admits not only advances in the fundamental understanding of operational aspects of quantum mechanics but also practical consequences. 
Quantum information processing protocols typically employ a pure state prepared in the standard form as its ``fuel''. 
For instance, quantum teleportation utilizes the maximally entangled state (called ``e-bit'') to realize the optimal performance. 
However, available quantum states are typically not provided in the desired form, mainly due to inevitable noise. This requires one to prepare such standard pure states from given noisy states.
This procedure is called \emph{resource distillation} and serves as a key routine in numerous settings, such as entanglement manipulation~\cite{Bennett1996concentrating}, fault-tolerant quantum computation~\cite{Bravyi2005universal}, and work extraction~\cite{Brandao2013resource,Horodecki2013fundamental}.
The generality of resource theories allows us to study the performance of the distillation process as a problem of free transformation. 
Indeed, distillation corresponds to a specific setting where we choose $\rho'$ as the desired pure state $\phi$.

The resource-theoretic framework can be applied to numerous physical settings to describe key quantities such as entanglement transformation under local operations and classical communication~\cite{Horodecki2009quantum}, quantum non-classicality~\cite{Yadin2018operational,Kwon2019nonclassicality} and non-Gaussianity~\cite{Genoni2010quantifying,Takagi2018convex,Albarelli2018resource} in continuous-variable systems, quantum thermodynamics~\cite{Horodecki2013fundamental,Brandao2013resource,Brandao15second_laws}, and non-Cliffordness (magicness) in fault-tolerant quantum computation~\cite{Veitch2014resource,Howard2017application}.  
In recent years, the unified understanding of the operational characterization of the individual quantum resources has been developed under the framework of \emph{general resource theories}, which keeps the generality of the choice of free objects and seeks the properties shared by all such quantum resources in the context of, e.g., advantages in discrimination tasks~\cite{Takagi2019operational,Takagi2019general,Uola2019quantifying,Regula2021operational}, resource quantification~\cite{Anshu2018quantifying,Regula2018convex,Liu2019resource,Takagi2019operational,Takagi2019general,Uola2019quantifying,Gour2019how,Gour2020dynamical_resources,Regula2021operational,Gonda2023monotones_general}, and resource distillation and dilution~\cite{Horodecki2013quantumness,Gour2017quantum,Liu2019one-shot,Regula2020benchmarking,Kuroiwa2020general,Fang2020nogo-state,Liu2020operational,Fang2020no-go,Regula2020oneshot,Regula2021fundamental,Regula2022functional,Regula2022tight_constraints,Takagi2022one-shot_yield-cost,Regula2022probabilistic}.
The main focus of this work is a specific instance of resource theory, which we introduce in the next subsection.
Nevertheless, having a background understanding and context in general quantum resource theories is helpful in interpreting and appreciating the results of individual theories. We also discuss extensions of our main results to general resource theories in Sec.~\ref{sec:general resource theories}.

\subsubsection{Quantum coherence and covariant operation}

Throughout this study, we consider the situation under the law of energy conservation.
Under this constraint, we cannot create coherence between energy eigenstates with different energies without any help, while it decoheres very easily.
We first clarify an {\it incoherent  state}, which has no coherence in it:

\bdf{[Incoherent state]
Consider a system with Hamiltonian $H=\sum_i E_i \Pi_i$ where $\Pi_i:=\sum_\alpha\ketbra{E_i,\alpha}{E_i,\alpha}$ is the projector onto the subspace of energy eigenstates with energy $E_i$. Here, $\ket{E_i,\alpha}$ is an energy eigenstate with energy $E_i$, and $\alpha$ distinguishes the degeneracy in the same energy.
A state $\rho$ is incoherent if this state is block-diagonal with respect to the energy eigenbasis: $\rho=\sum_i p_i\sigma_i$ with some probability distribution $\{p_i\}_i$ and state $\sigma_i$ such that $\Pi_i \sigma_i \Pi_i = \sigma_i$ for all $i$.
Equivalently, a state $\rho$ is incoherent if $\rho=e^{-iHt}\rho e^{iHt}$ holds for any $t\in \bbR$.
}

The equivalence can be confirmed as follows:
Let $\rho_{ij}:=\braket{i|\rho|j}$ be an off-diagonal element in the sense of block diagonalization, i.e., $\ket{i}$ and $\ket{j}$ are energy eigenstates with the energies $E_i$ and $E_j$ such that $E_i\neq E_j$.
Then, by expressing $\rho(t):=e^{-iHt}\rho e^{iHt}$, we have $\rho_{ij}(t)=e^{i(E_j-E_i)t}\rho_{ij}$.
Hence, $\rho_{ij}(t)=\rho_{ij}$ is equivalent to $\rho_{ij}=0$ for $E_i\neq E_j$.

Since a coherent state cannot be prepared from an incoherent state under energy conservation, coherence among energy eigenstates with different energies is considered as a precious resource in this setting, while incoherent states can be regarded as free states which we can freely use and waste.

The class of possible operations under the law of energy conservation is characterized by {\it covariant operations}.
We consider a state transformation from system $S$ to system $S'$, whose Hamiltonians are respectively $H_S$ and $H_{S'}$.
Let $\rho$ and $\rho'$ be states in $S$ and $S'$.
We first present one definition of covariant operations whose physical picture is most transparent and then provide equivalent definitions that are more axiomatic.

\bdf{[Covariant operation]
An operation $\Lambda: S\to S'$ is a covariant operation if the following relation is satisfied:
There exist auxiliary systems $A$ and $A'$ with Hamiltonians $H_A$ and $H_{A'}$ satisfying $S\otimes A=S'\otimes A'$ such that the operation $\Lambda$ can be expressed as
\eq{
\Lambda(\rho)=\Tr_{A'}[U(\rho\otimes \eta)U^\dagger],
}
where $U$ is an energy-conserving unitary satisfying
\eq{
U(H_S\otimes I_A+I_S\otimes H_A)U^\dagger =H_{S'}\otimes I_{A'}+I_{S'}\otimes H_{A'},
}
with the identity operator $I$, and $\eta$ is an incoherent state in $A$. 
}

The above definition manifests the fact that a covariant operation is implementable under the law of energy conservation with an incoherent state.

We remark that there are several equivalent characterizations of covariant operations:
The equivalence of these definitions is proven in, e.g., Refs.~\cite{Keyl1999optimal,Mar20}.

\begin{pro}
An operation $\Lambda$: $S\to S'$ is a covariant operation if and only if
\eq{
\Lambda(e^{-iH_St}\rho e^{iH_St})=e^{-iH_{S'}t}\Lambda(\rho)e^{iH_{S'}t}
}
holds for any $\rho$ and any $t\in \bbR$.
\end{pro}

\begin{pro}\label{pro:covariant_app}
An operation $\Lambda$: $S\to S'$ is a covariant operation if and only if the following condition is satisfied:
For any auxiliary system $B$ with Hamiltonian $H_B$, if a state $\tau$ on $SB$ is an incoherent state with respect to $H_S\otimes I_B+I_S\otimes H_B$, then the final state $\Lambda\otimes \calI_B(\tau)$  is also an incoherent state with respect to $H_{S'}\otimes I_B+I_{S'}\otimes H_B$.
Here, $\Lambda\otimes \calI_B$ is an operation that applies $\Lambda$ on system $S$ and leave system $B$ as it is.
\end{pro}

The above operational setting provides a resource theory with the set $\bbF$ of free states being incoherent states and the set $\bbO$ of free operations being covariant operations. 
This was introduced under the name of the resource theory of asymmetry~\cite{GS08}, as the coherence is manifested by the resource that causes the asymmetry under time translation $\{e^{-iHt}\}_{t\in\mathbb{R}}$, which constructs a unitary representation of the group $\mathbb{R}$.
Therefore, we also use the word ``asymmetry'' interchangeably to refer to coherence.
Although the framework of resource theory of asymmetry encompasses the general group $G$, in this manuscript we refer to the case $G=\mathbb{R}$ with time translation as its unitary representation unless otherwise stated.
For instance, we say ``asymmetry measure'' to refer to a resource measure defined for the resource theory of asymmetry in the above sense. 

We also remark on the potential confusion regarding the use of the word ``coherence''.
In the context of resource theories, another standard framework to describe quantum superposition is to look at off-diagonal terms with respect to a given orthonormal basis $\{\ket{i}\}_i$~\cite{Baumgratz2014quantifying}, where there is no concept of energy or time translation. 
The coherence characterized in this framework is known as \emph{speakable} coherence, while coherence relevant in this work is known as \emph{unspeakable} coherence~\cite{Marvian2016howto}.
The former quantity is rooted in the computational notion where we are given a fixed computational basis, while the latter is relevant to the physical system equipped with a certain Hamiltonian, making it a relevant quantity for quantum clock and work extraction in the quantum thermodynamic setting. 
The properties of these two different notions of coherence are significantly different --- we indeed see that our main results regarding unbounded amplification of coherence never apply to the speakable coherence, as we discuss in Sec.~\ref{subsec:restriction measure}.

\subsection{Coherent modes}

To formally state our result, let us introduce the notion of \emph{coherent modes}. 
The idea of coherent modes stems from the observation that the difference in the energy levels to which coherence is attributed plays a crucial role in characterizing the coherence transformation. 
Classifying coherence in terms of energy levels can be formalized by the \emph{modes of asymmetry} introduced in Ref.~\cite{Marvian2014modes}. 
\begin{dfn}[Modes of asymmetry~\cite{Marvian2014modes}] \label{def:mode of asymmetry}
The modes of asymmetry is defined as the set of a mode with non-zero coherence: 
\begin{equation}
\calD(\rho):=\lset \Di_{ij} \sbar \Di_{ij}=E_i-E_j,\ (i,j)\in \calM(\rho)\rset,
\end{equation}
where $\calM(\rho):=\{ (i,j)| \braket{E_i,\alpha|\rho|E_j,\beta}\neq 0\}$ is the set of integer pairs $(i,j)$ such that the off-diagonal element of $\rho$ with respect to energy eigenstates with energies $E_i$ and $E_j$ is non-zero. 
\end{dfn}

As we shall show in the remainder, extremely small but non-zero coherence can be amplified arbitrarily, while exactly zero coherence never becomes non-zero coherence.
Therefore, we need to determine whether state $\rho$ has coherence on the modes where $\rho'$ also has coherence.
To describe the presence and the absence of coherence on a mode, we define a set of \emph{resonant coherent modes} in state $\rho$ on a system equipped with energy levels $\{ E_i\}_i$, which is an extension of the mode of asymmetry~\cite{Marvian2014modes}.

\bdf{[Set of resonant coherent modes]\label{def:set of resonant coherent modes}
Denoting by $\Di_{ij}:=E_i-E_j$, we define a set of resonant coherent modes $\calC(\rho)$ as a linear combination of non-zero coherent mode of $\rho$ with integer coefficients:
\eq{
\calC(\rho):=\lset x \sbar x=\sum_{(i,j)\in \calM} n_{ij}\Di _{ij} : n_{ij}\in \bbZ\rset .
}
}

Here we consider a linear combination of coherent mode because if we have coherent modes with energy difference $1$ and $\sqrt{2}$, then we can create a coherent mode with energy difference $1+\sqrt{2}$ through a covariant operation in both the asymptotic-marginal and correlated-catalytic transformation~\cite{TS22}.
The relation $\calC(\rho')\subseteq \calC(\rho)$ means that all resonant coherent modes of $\rho'$ are also those of $\rho$.

A simple but important fact on $\calC(\rho)$ is that $\rho$ and its tensor-product state $\rho^{\otimes n}$ has the same set of resonant coherent modes:
\eq{
\calC(\rho^{\otimes n})=\calC(\rho).
}


\section{Asymptotic transformation}\label{sec:asymptotic}

\subsection{Preliminaries} \label{subsec:asymptotic preliminaries}

Before proceeding to our main arguments, we first present general frameworks of resource theories not restricted to quantum coherence, and then proceed to quantum coherence.
One of the effective approaches to studying free transformation is to leverage the tools and ideas from Shannon theory, aiming to transform many copies of the initial state into many copies of the target state. 
We then study the rate of the number of copies of the initial and final states. This framework is generally called \emph{asymptotic transformation}.  

There are several approaches to quantifying the performance of asymptotic transformation. 
One of the ways is to consider the maximum transformation rate at which a free operation can transform many copies of $\rho$ into a state that approaches many copies of the target state $\rho'$. 
Specifically, we call $r$ an achievable rate if for any $\ep>0$ there exists a series $\{\Lambda_n\}_n$ of free operations (i.e., $\Lambda_n\in\bbO$ for all $n$) such that $\lim_{n\to\infty}\|\Lambda_n(\rho^{\otimes n})-{\rho'}^{\otimes \lfloor rn \rfloor}\|_1 <\ep$. The \emph{asymptotic transformation rate} $R(\rho\to\rho')$ is the supremum of the achievable rates given by
\bal
 R(\rho\to\rho')&:=\sup\lset r \sbar \lim_{n\to\infty}\|\Lambda_n(\rho^{\otimes n})-{\rho'}^{\otimes \lfloor rn\rfloor}\|_1 = 0,\ \Lambda_n\in\bbO:S^{\otimes n}\to {S'}^{\otimes \lfloor rn \rfloor} \rset .
\eal
In particular, when the target state $\rho'$ is a pure state $\phi$, the rate $R(\rho\to\phi)$ is customarily called \emph{asymptotic distillation rate} with the target state $\phi$.
We note that there may exist a finite error in the final state for finite $n$, while it should vanish in the asymptotic limit.

The asymptotic transformation rate is relevant when one would like to prepare uncorrelated copies of the target state $\rho'$. 
Another operational setting is when multiparties are separate apart and each of them would like to obtain a state close to the target state $\rho'$. A reasonable goal in this setting is to obtain a state whose local marginal states are close to the target state while distributing the good local state to as many parties as possible. 
The transformation rate suitable for characterizing such a setting was previously studied~\cite{Fer23,Ganardi2023catalytic}. 
Here, we call it \emph{asymptotic marginal transformation rate} and formally define it as follows.

\begin{dfn}[Asymptotic marginal transformation rate]
Let $\rho$ and $\rho'$ be states on systems $S$ and $S'$. We say that asymptotic marginal transformation rate $r$ is achievable if there is a series $\{\Lambda_n\}_{n}$ of free operations with $\Lambda_n:S^{\otimes n}\to S'^{\otimes \lfloor rn \rfloor}$ such that $\lim_{n\to\infty}\max_i \|\Tr_{\backslash i}\Lambda_n(\rho^{\otimes n})-\rho'\|_1=0$. 
Here, $\Tr_{\backslash i}$ represents a partial trace over subsystems except for the $i$\,th one. 
The \emph{asymptotic marginal transformation rate} $\tilde R(\rho\to\rho')$ is the supremum over the achievable rates, i.e., 
\bal
 \tilde R(\rho\to\rho')&:=\sup\lset r \sbar \lim_{n\to\infty}\max_i \|\Tr_{\backslash i}\Lambda_n(\rho^{\otimes n})-\rho'\|_1 = 0,\ \Lambda_n:S^{\otimes n}\to {S'}^{\otimes \lfloor rn \rfloor}\in\bbO \rset .
\eal
\end{dfn}

When $\rho'$ is a pure state $\phi$, we particularly call $\tilde R(\rho\to\phi)$ \emph{asymptotic marginal distillation rate} with the target state $\phi$.
Unlike the standard asymptotic transformation, the asymptotic marginal transformation does not require the final state to approach a product of the target state at the infinite-copy limit, allowing for some correlation among the subsystem in the final state. 
However, if $\lfloor rn\rfloor$ copies are used separately and do not interact with each other, the final state is indistinguishable from the case of an exact transformation.

The contractivity of the trace distance under the partial trace ensures that $\|\Lambda_n(\rho^{\otimes n})-{\rho'}^{\otimes \lfloor rn \rfloor}\|_1<\varepsilon$ implies $\|\Tr_{\backslash i}\Lambda_n(\rho^{\otimes n})-\rho'\|_1<\varepsilon$, leading to $R(\rho\to\rho')\leq \tilde R(\rho\to\rho')$. 
The previous study~\cite{Fer23} found that the asymptotic marginal transformation is explicitly upper bounded as 
\bal
 R(\rho\to\rho')\leq \tilde R(\rho\to\rho')\leq \frac{G(\rho)}{G(\rho')},
\eal
where $G$ is a resource measure that is superadditive, tensor-product additive, and lower semicontinuous. 
In the case of the theory of entanglement, a much stronger claim was shown.
Let $\ket{\Phi}=\frac{1}{\sqrt{2}}(\ket{00}+\ket{11})$ be the maximally entangled state.
If $\rho$ is distillable (i.e., $R(\rho\to\Phi)>0$), the above two rates coincide; $R(\rho\to\Phi)=\tilde R(\rho\to\Phi)$~\cite{Ganardi2023catalytic}.

Our result also involves a stronger notion than the asymptotic marginal transformation, which we call \emph{asymptotic exact marginal transformation}.
In the asymptotic transformation, one typically allows non-zero errors that vanish asymptotically at the infinite-copy limit.
This can be understood via the non-asymptotic transformation with non-zero error. 
Let $\rho$ and $\rho'$ be states on system $S$ and $S'$. 
Then we define the asymptotic marginal transformation rate with the target state $\rho'$ by
\begin{equation}
 \tilde R(\rho\to\rho') := \lim_{\ep\to 0} \lim_{n\to\infty}\sup \lset r  \sbar \max_i \|\Tr_{\backslash i}[\Lambda_{n}(\rho^{\otimes n})]-\rho'\|_1<\varepsilon, \ \Lambda_n:S^{\otimes n}\to {S'}^{\otimes \lfloor rn \rfloor}\in \bbO \rset.
\end{equation}
As is explicit in the above formulation, the asymptotic marginal transformation admits non-zero errors for every $n$ as long as it asymptotically vanishes at the limit $n\to\infty$. 
On the other hand, we can also define the \emph{asymptotic exact marginal transformation} and its rate as follows:
\begin{equation}
 \tilde R^{0}(\rho\to\rho') := \lim_{n\to\infty}\sup \lset r  \sbar \forall i \ \Tr_{\backslash i}[\Lambda_{n}(\rho^{\otimes n})]=\rho' , \  \Lambda_n:S^{\otimes n}\to {S'}^{\otimes \lfloor rn \rfloor}\in \bbO \rset.
\end{equation}
As seen from the definition, the asymptotic exact marginal transformation does not allow any error, which is a much more stringent requirement than the asymptotic marginal transformation.
In fact, since every sequence of operation $\{\Lambda_n\}_n$ that achieves the asymptotic marginal exact transformation clearly realizes the transformation whose error vanishes at the limit of $n\to\infty$, we always have $\tilde R^0(\rho\to\rho')\leq \tilde R(\rho\to\rho')$.

We note that despite the clear relation between two asymptotic rates $R(\rho\to\rho')\leq \tilde R(\rho\to\rho')$, there is no simple relation between the asymptotic rate $R$ and asymptotic exact marginal rate $\tilde R^0$.
The asymptotic exact marginal transformation rate can similarly be introduced for the standard asymptotic transformation, and it was extensively studied in the context of zero-error distillable entanglement and entanglement cost~\cite{Audenaert2003entanglement,Regula2019one-shot,Wang2020cost}.

\bigskip

In the theory of coherence, the performance of the standard asymptotic transformation rate was studied in the context of coherence distillation. 
Notably, Ref.~\cite{Mar20} showed a fundamental restriction in the coherence distillation --- every full-rank state has a zero asymptotic distillation rate. 
\bthm{[Coherence distillation is impossible~\cite{Mar20}]\lb{t:distill-impossible}
Consider the asymptotic transformation with covariant operations. 
For an arbitrary full-rank state $\rho$ and a target pure coherent state $\phi$, $R(\rho\to\phi)=0$ holds. 
}

This result appears to suggest that obtaining high-quality coherent bits would come with a fundamental difficulty.  
However, we show that the distillation rate behaves in a dramatically different manner if we consider the asymptotic marginal transformation. 
Namely, we prove that these two rates take two opposite extremes, and the difference between these two becomes unbounded, realizing $R(\rho\to\phi)=0$ and $\tilde R(\rho\to\phi)=\infty$ for almost all coherent states $\rho$. 
To the best of our knowledge, this is the first example of the resource theory for which these two rates do not coincide. 
We also show that the coherence manipulation is free from the severe restriction of the exact transformation when it comes to the asymptotic marginal transformation. The asymptotic exact marginal rate $\tilde R^0$ also diverges for most coherent states. 

\subsection{Arbitrary coherence amplification: asymptotic marginal transformation (Theorem~\ref{thm:asymptotic})}

Now we are in a position to prove our first main result Theorem~\ref{thm:asymptotic} in the main text.
We break down the claim of Theorem~\ref{thm:asymptotic} on the asymptotic marginal transformation into two parts: the diverging rate for $\mathcal{C}(\rho')\subseteq\mathcal{C}(\rho)$ (discussed in this subsection) and vanishing rate for $\mathcal{C}(\rho')\nsubseteq\mathcal{C}(\rho)$ (discussed in Sec.~\ref{subsec:asymptototic no go appendix}).
We will see that the resonant coherent modes introduced above play an essential role in dividing these two regimes and completely characterizes the power of asymptotic marginal transformation.

We begin with the case when the infinite rate can be realized. 
We also show that the strength of correlation can be made arbitrarily small, which makes the asymptotic marginal transformation even more operationally relevant.

\begin{thm}[The first part of Theorem~\ref{thm:asymptotic} in the main text]\label{thm:asymptotic go}
    For arbitrary states $\rho$ and $\rho'$, $\tilde R(\rho\to\rho')$ diverges if $\mathcal{C}(\rho')\subseteq\mathcal{C}(\rho)$. 
    Moreover, the correlation between one subsystem and the others can be made arbitrarily small, i.e., for any $\ep>0$ the final state $\tau$ satisfies $\| \tau-\rho'\otimes \Tr_{i}\tau\|_1<\ep$ for any $i$-th copy. 
\end{thm}

In fact, we show an even stronger claim, which shows that asymptotic exact marginal transformation rates also diverge for almost all transformations.

\begin{thm}\label{thm:asymptotic exact go}
Suppose $\rho$ and $\rho'$ are arbitrary states such that $\mathcal{C}(\rho')\subseteq\mathcal{C}(\rho)$ and $\rho'$ is full rank.
Then, the asymptotic exact marginal transformation rate $\tilde R^0(\rho\to \rho')$ diverges.
Moreover, the correlation between one subsystem and the others can be made arbitrarily small, i.e., for any $\ep>0$ the final state $\tau$ satisfies $\| \tau-\rho'\otimes \Tr_{i}\tau\|_1<\ep$ for any $i$-th copy. 
\end{thm}

We remark that Theorem~\ref{thm:asymptotic go} is a direct consequence of Theorem~\ref{thm:asymptotic exact go} as shown in the following.

\begin{proof}[Proof of Theorem~\ref{thm:asymptotic go}]
Theorem~\ref{thm:asymptotic go} can be derived from Theorem~\ref{thm:asymptotic exact go} by noting that full-rank states are dense in the state space and thus every state has a full-rank state in its arbitrary neighborhood.
Let $\rho'$ be an arbitrary state in $S'$ satisfying $\mathcal{C}(\rho')\subseteq\mathcal{C}(\rho)$. 
For a given $\delta>0$, there exists a full-rank state $\tilde\rho$ in $S'$ such that $\|\rho'-\tilde\rho\|_1<\delta$. Theorem~\ref{thm:asymptotic exact go} ensures that for every $r>0$, there exists an integer $n$ and a covariant operation $\Lambda:S^{\otimes n}\to {S'}^{\otimes \lfloor rn \rfloor}$ such that $\Tr_{\backslash i}\Lambda(\rho^{\otimes n})=\tilde\rho$ for all $i$. 
This ensures that 
\bal
\|\Tr_{\backslash i}\Lambda(\rho^{\otimes n})-\rho'\|_1&\leq \|\Tr_{\backslash i}\Lambda(\rho^{\otimes n})-\tilde\rho\|_1+\|\rho'-\tilde\rho\|_1\\
&<\delta
\eal
for every $i$, which shows that $R(\rho\to\rho')$ diverges.
It is also guaranteed by Theorem~\ref{thm:asymptotic exact go} that the final state $\tau=\Lambda(\rho^{\otimes n})$ satisfies $\|\tau-\rho'\otimes \Tr_{\backslash i}\tau\|_1<\varepsilon$ for all $i$.
\end{proof}

Therefore, we focus on proving Theorem~\ref{thm:asymptotic exact go}.
To this end, we introduce several results needed for the proof.

One of the key observations in proving our theorems is that any state conversion is possible by a covariant operation with the help of a class of auxiliary states called \emph{marginal catalysts}.

\begin{dfn}[Marginal-catalytic transformation]\label{def:marginal catalytic}
We say that a state $\xi$ in system $S$ is convertible to a state $\rho'$ in the system $S'$ by a marginal-catalytic free transformation if there exists finite-dimensional catalytic systems $C_1,\dots, C_M$ with states $c_1,\dots,c_M$ and a free operation $\Lambda: S\otimes C_1\otimes\dots\otimes C_M\to S'\otimes C_1\otimes\dots\otimes C_M\in\bbO$ such that
\eq{
\tau=\Lambda(\xi\otimes c_1\otimes\dots\otimes c_M), \hspace{10pt} \Tr_{\backslash S}\tau=\rho', \hspace{10pt} \Tr_{\backslash C_i}\tau=c_i\ \forall i.
}
\end{dfn}
The marginal-catalytic transformation was first introduced in the context of quantum thermodynamics~\cite{Lostaglio2015stochastic}, and it was recently shown that an arbitrary state transformation is possible by a covariant operation with a marginal catalyst~\cite{TS22}.
The precise statement is as follows:

\begin{lm}[Theorem 2 in Ref.~\cite{TS22}]\label{t:marginal-catalyst}
For any two quantum states $\xi$ and $\rho'$, and for any accuracy $\ep>0$,  there exists a set of two-level catalytic systems $C_1,\ldots , C_N$ with full-rank qubit mixed states $c_1,\ldots , c_N$ and a covariant operation $\Lambda$: $S\otimes C_1\otimes \cdots \otimes C_N\to S'\otimes C_1\otimes \cdots \otimes C_N$  such that $\Lambda(\xi\otimes c_1\otimes \cdots \otimes c_N)=\tau$ with $\| \Tr_{C_1,\ldots , C_N}[\tau]-\rho'\|_1<\ep$ and $\Tr_{\backslash C_i}[\tau]=c_i$ for all $i=1,\ldots , N$. 
\end{lm}

Note that although the above statement is on the marginal-catalytic transformations with vanishing error, as demonstrated soon after (Proposition.~\ref{t:marginal-catalyst-modify}) this can be strengthened into the form of the exact marginal-catalytic transformations.

The framework of marginal-catalytic transformations comes with a less clear operational meaning compared to the correlated catalyst, as the final catalyst cannot be reused indefinitely due to the correlation among catalytic subsystems.  
Nevertheless, we show that the marginal-catalytic covariant transformation constructed in Ref.~\cite{TS22} serves as an effective subroutine in our protocols.

The key property of marginal catalysts that lends themselves to useful subroutines is that, although the marginal catalysts cannot be reused indefinitely for state transformation, they can still give some operational advantage. 
It was shown that they are \emph{partially} reusable, meaning that the catalyst allows a larger number of transformations than the number of sets of catalysts provided.  
For brevity, we denote a set $C_1,\ldots , C_N$ of catalytic systems and their initial states $c_1,\ldots , c_N$ by $\bsC$ and $\bsc$, respectively.

\begin{lm} [Supplemental Material of Ref.~\cite{TS22}]\label{t:recomb}
Let $\xi$ and $\rho'$ be arbitrary states in systems $S$ and $S'$. 
For any accuracy $\ep>0$, let $\bsC$ and $\bsc$ be the sets of catalytic systems and states ensured by Lemma~\ref{t:marginal-catalyst}.
Then, $N^k$ sets of catalysts $\bsc^{\otimes N^k}$ in $\bsC^{\otimes N^k}$ can realize $(k+1)N^k$ marginal-catalytic transformation from $\xi$ to $\rho'$, i.e., there is a covariant operation $\Lambda$: $S^{\otimes (k+1)N^k}\otimes \bsC^{\otimes N^k}\to S'^{\otimes (k+1)N^k}\otimes \bsC^{\otimes N^k}$ such that
\eq{
\Lambda(\xi^{\otimes (k+1)N^k}\otimes \bsc^{\otimes N^k})=\tau, \hspace{10pt} \|\Tr_{\backslash S_j}[\tau] -\rho'\|_1<\ep, \hspace{10pt} \Tr_{\backslash C_{ij}}[\tau]=c_i\ \forall i
}
for every $1\leq j\leq (k+1)N^k$.
Here, $S_j$ denotes the $j$\,th main system, and $C_{ij}$ refers to the $i$\,th catalyst in the $j$\,th copy.
\end{lm}

The idea is to reuse the marginal catalysts by recombining them in a way that the correlation present in the final state does not affect the next round of transformation. 
Below we present a way of such a recombination.
We label $N^k$ copies of catalysts $C_j$ by a tuple of $k$ integers $(n_1,n_2,\ldots , n_k)$ with $n_i\in \{1,2,\ldots , N\}$.
In the first step, we coordinate catalysts into $N^k$ groups of $C_1,\ldots , C_N$ such that catalysts in the same group have the same label $(n_1,n_2,\ldots , n_k)$.
Using these $N^k$ sets of catalysts, we have $N^k$ conversions in this step.
In the $l$-th step ($2\leq l\leq k+1$), we coordinate catalysts into $N^k$ groups of $C_1,\ldots , C_N$ in the following manner:
$N$ catalysts in the same group has the same labels $(n_1,n_2,\ldots , n_k)$ except for $n_l$, and there exists an integer $g$ such that the label $n_l$ with catalyst $C_j$ is written as $n_l=g+j \mod N$ for all $C_j$.
Using these $N^k$ sets of catalysts, in each step we have $N^k$ conversions.
It is easy to confirm that in any group all the catalysts have no correlation before the catalytic transformation.
Through the whole procedure, we have $(k+1)N^k$ conversions.
We refer interested readers to the first section of the Supplemental Material of Ref.~\cite{TS22} for further details.

\medskip

We also introduce a useful lemma, which connects approximate transformations to exact transformations.
The essential idea was shown by Wilming~\cite{Wilming2022correlations} for the catalytic entropy conjecture, and following this we present a statement in a general form.

\blm{[Ref.~\cite{Wilming2022correlations}]\label{t:exact}
We consider a quantum system equipped with the trace distance $d(\rho,\sigma):=\frac12 \|\rho-\sigma\|_1$.
Given a sequence of convex closed sets of quantum states $\{S_n\}_{n=1}^\infty$ satisfying $S_n\subseteq S_{n+1}$,
let $V$ be a set of states such that for any $\ep>0$ and any $\sigma \in V$ there exist a sufficiently large $N$ and a state $\eta\in S_N$ such that $d(\sigma,\eta)<\ep$. If $\kappa$ is an interior state of $V$ in terms of distance $d$, then there exists an integer $n$ such that $\kappa\in S_n$.
}

Let $S_n$ be the set of exactly convertible states with parameter $n$ (e.g., the number of copies and the size of a catalyst), and let $V$ be the set of approximately convertible states from $\rho$.
Lemma~\ref{t:exact} suggests that if we can convert $\rho$ to $\kappa$ approximately and $\kappa$ is an interior state of $V$, then we can convert $\rho$ to $\kappa$ exactly.
The condition of convexity is fulfilled if a classical mixture is a free operation in the resource theory in consideration.
This fact enables us to interpret results on approximate transformations to exact transformations.

\begin{proof}[Proof of Lemma~\ref{t:exact}]
We prove it by contradiction.
Suppose contrarily that $\kappa\in V$ is an interior state while it does not belong to $S_\infty:=\lim_{n\to \infty}S_n$.

Consider an open ball $B_\delta=\lset \rho \sbar \|\rho-\kappa\|_1 \leq \delta,\ \rho\geq 0,\ \Tr(\rho)=1 \rset$ with its center $\kappa$ and radius $\delta$ such that $B_\delta\subseteq V$.
By definition, there exists $n$ such that for any $\eta\in B_\delta$, $S_n$ has a state $\xi(\eta) \in S_n$ satisfying $\|\xi(\eta)-\eta\|_1 <\delta/2$.
Since $S_n$ is convex and $\kappa \notin S_n$, there uniquely exists a state $a\in S_\infty$ closest to $\kappa$.
Let $\rho(t) := -(t-1)a + t\kappa,\  t\geq 1$ be the set of states on the ray from $\kappa$ along the line going through $a$ and $\kappa$. 
Since $\Tr(\rho(t))=1$ for every $t$, $\rho(t)$ is a quantum state if and only if $\rho(t)\geq 0$.
We also have $\|\rho(t)-\kappa\|_1 = (t-1)\|a-\kappa\|_1$, which is continuous and increasing with $t$.
Therefore, for every $\delta'<\delta$, there exists $t'\geq 1$ such that $\|\rho(t')-\kappa\|_1=\delta'$ and $\rho(t')\in B_\delta$.
Taking $\delta'=2\delta/3$, we get 
\begin{equation}
\|\rho(t')-a\|=t'\|a-\kappa\|_1\geq (t'-1)\|a-\kappa\|_1=\|\rho(t')-\kappa\|_1 = 2\delta/3.
\end{equation}
We also note that 
\begin{equation}
\min_{b\in S_n}\|\rho(t')-b\|=\min_{b\in S_n}\|-(t'-1)a+t'\kappa - b \|_1 = t'\min_{b\in S_n}\left\|\kappa - \left[(1-{t'}^{-1})a 
 + {t'}^{-1} b\right]\right\|_1 = t'\|\kappa - a\|_1
\end{equation}
where in the last equality, we used the fact that $(1-{t'}^{-1})a + {t'}^{-1}b\in S_n$ due to the convexity of $S_n$ and the assumption that $a$ is the closest state to $\kappa$ and thus $b=a$ achieves the minimum. 
Together with the assumption of $S_n$ that $\min_{b\in S_n}\|\rho(t')-b\|\leq \delta/2$, we obtain 
\begin{equation}
    \delta/2 \geq \min_{b\in S_n}\|\rho(t')-b\| = \|\rho(t')-a\| \geq \|\rho(t')-\kappa\|_1 = 2\delta/3
\end{equation}
which is a contradiction.

\end{proof}

Applying Lemma~\ref{t:exact} to Lemma~\ref{t:marginal-catalyst}, we have the following proposition, showing that, if the target state is full rank, the marginal transformation can be made exact.  

\bpro{\label{t:marginal-catalyst-modify}
For any quantum state $\xi$ and a full-rank state $\rho'$, there exists a set of two-level catalytic systems $C_1,\ldots , C_N$ with full-rank mixed states $c_1,\ldots , c_N$ and a covariant operation $\Lambda$ such that $\Lambda(\xi\otimes c_1\otimes \cdots \otimes c_N)=\tau$ with $\Tr_{C_1,\ldots , C_N}[\tau]=\rho'$ and $\Tr_{\backslash C_i}[\tau]=c_i$ for all $i=1,\ldots , N$.
}

We also show that an arbitrarily good qubit coherent state can be prepared from finite copies of a general state $\rho$ as long as it contains the mode of asymmetry for the qubit coherent state (recall Definition~\ref{def:mode of asymmetry}).

\begin{lm}\label{t:distill}
    Let $\rho$ be an arbitrary state and $\sigma$ be an arbitrary two-level state such that $\mathcal{D}(\sigma)\subseteq\mathcal{D}(\rho)$. Then, for every $\varepsilon>0$, there exists a positive integer $n$ and a covariant operation $\Lambda$ such that $\|\Lambda(\rho^{\otimes n})-\sigma\|_1<\varepsilon$.
\end{lm}
\begin{proof}
   For every $\Di \in \mathcal{D}(\rho)$, one can transform $\rho$ to a weakly coherent qubit state $\eta$ with $\mathcal{D}(\eta) = \{0,\pm\Di\}$ by a covariant operation using the protocol in Ref.~\cite[Lemma 11 in Supplemental Material]{TS22}.
   Let $\Lambda_1$ be this covariant operation such that $\Lambda_1(\rho)=\eta$.
   As pointed out in Ref.~\cite{Mar20}, the protocol introduced in Ref.~\cite{Cirac1999optimal} allows one to transform many weakly coherent qubit states to one copy of a good coherent state by a covariant operation. That is, for every $\varepsilon'>0$, there is a positive integer $m$ and a covariant operation $\Lambda_2$ such that $\|\Lambda_2(\eta^{\otimes m})-\dm{+}\|_1<\varepsilon'$ where $\ket{+}:=\frac{1}{\sqrt{2}}(\ket{0}+\ket{1})$ is the maximally coherent state with $\Di\in\mathcal{D}(\dm{+})$.

   We now employ the fact that every full-rank qubit state $\sigma$ can be prepared exactly by a covariant operation from finite copies of $\dm{+}$ defined in the same system. This is because the coherence cost $1/R(\dm{+}\to\sigma)$ is proportional to the ratio of the quantum Fisher information of $\sigma$ to that of $\ket{+}$~\cite{Marvian2022operational}, and $\ket{+}$ has the largest quantum Fisher information in a two-level system, particularly ensuring that the coherence cost is upper bounded by 1.   
   Therefore, for every $\delta>0$ there is an integer $k'$ and a covariant operation $\tilde\Lambda_3$ such that $\|\tilde\Lambda_3(\dm{+}^{\otimes k'})-\sigma^{\otimes k'}\|_1<\delta$. 
   Taking the partial trace other than the first subsystem and noting the data-processing inequality of the trace norm, we get $\|\Lambda_3'(\dm{+}^{\otimes k'})-\sigma\|_1<\delta$ where $\Lambda_3'=\Tr_{\backslash 1}\circ\tilde\Lambda_3$ is also a covariant operation.  
   We then note that this applies to an arbitrary qubit state $\sigma$, which ensures the existence of the exact transformation that prepares $\sigma$ because of Lemma~\ref{t:exact}.
   We let $\Lambda_3$ be such a covariant operation satisfying $\Lambda_3(\dm{+}^{\otimes k})=\sigma$ for some integer $k$.
   
   Then, 
   \bal       
   \|\Lambda_3\circ\Lambda_2^{\otimes k}\circ\Lambda_1^{\otimes km}(\rho^{\otimes km}) - \sigma\|_1 &\leq \|\Lambda_3\circ\Lambda_2^{\otimes k}(\eta^{\otimes km}) - \Lambda_3(\dm{+}^{\otimes k})\|_1 \\
       &\leq \|\left[\Lambda_2(\eta^{\otimes m})\right]^{\otimes k} - \dm{+}^{\otimes k}\|_1\\
       &< k\varepsilon' 
       \label{eq:preparing qubit state}
   \eal
   where in the last inequality we used $\|\Lambda_2(\eta^{\otimes m})-\dm{+}\|_1\leq \varepsilon'$ and the following general inequality satisfying for all states $\rho_1$ and $\rho_2$ and every integer $n$:
   \bal
    \|\rho_1^{\otimes n}-\rho_2^{\otimes n}\|_1 &\leq \|\rho_1^{\otimes n} - \rho_2\otimes \rho_1^{\otimes n-1}\|_1 + \|\rho_2\otimes \rho_1^{\otimes n-1}-\rho_2^{\otimes 2}\otimes \rho_1^{\otimes n-2}\|_1 + \dots + \|\rho_2^{\otimes n-1}\otimes \rho_1 - \rho_2^{\otimes n}\|_1 \\
    &\leq n \|\rho_1-\rho_2\|_1.
   \eal
 Since $\ep'$ in \eqref{eq:preparing qubit state} can be taken as small as one wishes, the target error $\ep$ can be realized by choosing $\ep'=\ep/k$. 
Noting that $\Lambda:=\Lambda_3\circ\Lambda_2^{\otimes k}\circ\Lambda_1^{\otimes km}$ is also a covariant operation concludes the proof. 
\end{proof}

Using Lemma~\ref{t:exact}, we can improve Lemma~\ref{t:distill} as the exact preparation of arbitrary full-rank final states.

\begin{lm}\label{t:distill-exact}
    Let $\rho$ be an arbitrary state and $\sigma$ be an arbitrary two-level full-rank state such that $\mathcal{D}(\sigma)\subseteq\mathcal{D}(\rho)$. Then, there exists a positive integer $n$ and a covariant operation $\Lambda$ such that $\Lambda(\rho^{\otimes n})=\sigma$.
\end{lm}

We are now in a position to present the proof of Theorem~\ref{thm:asymptotic exact go}.

\begin{proof}[\it Proof of Theorem~\ref{thm:asymptotic exact go}]
We first construct an asymptotic conversion protocol whose transformation rate is larger than any integer $R$ without taking into account the small correlation condition, and then demonstrate how to suppress correlation.

Let $\rho$ be a state in $S$ and $\rho'$ be a full-rank state in $S'$ such that $\calC(\rho')\subseteq \calC(\rho)$.
We aim to construct the catalysts $c_1,\dots, c_N$ by a covariant operation from multiple (but finite) copies of the initial state $\rho$. 
Lemma~\ref{t:distill-exact} ensures that there exists $\{\mu_i\}_{i=1}^N$ such that $\rho^{\otimes \mu_i}$ is convertible to $c_i$ exactly by a covariant operation.
We write $\mu:=\sum_{i=1}^N\mu_i$.
We remark that the exact preparation of the catalysts here is essential to avoid the errors from accumulating during the multiple uses of the prepared catalysts $c_1,\dots, c_N$.

We now construct the desired covariant operation as follows:
We start with $\mu N^k$ copies of $\rho$.
We first convert $\mu N^k$ copies of $\rho$ into $N^k$ copies of $c_1\otimes \cdots \otimes c_N$ by the above protocol.
Then, using the partial reusability of the marginal catalysts shown in Lemma~\ref{t:recomb} and the exact marginal-catalytic covariant transformation ensured by Proposition~\ref{t:marginal-catalyst-modify}, a set of free states $\xi^{\otimes (k+1)N^k}$ can be converted into state $\Sigma$ on $S'^{\otimes (k+1)N^k}$ such that its reduced state to any single copy $i=1,\ldots , (k+1)N^k$ is equal to $\rho'$ exactly ($\Tr_{\bcs i}[\Sigma]=\rho'$ for all $1\leq i\leq (k+1)N^k$).
Overall, this protocol realizes the covariant operation that transforms $\rho^{\otimes n}$ to $m$ copies of $\rho'$ with $n:=\mu N^k$ and $m:=(k+1)N^k$ in the sense of the marginal asymptotic conversion. 

Since 
\eq{
\frac{m}{n}=\frac{(k+1)N^k}{\mu N^k}=\frac{k+1}{\mu}
}
can become arbitrarily large by taking a sufficiently large $k$, we get that any transformation rate $R$ is an achievable asymptotic exact marginal transformation rate.

We now demonstrate how to ensure that the correlation between a copy and the remainder of copies is less than $\ep$ in the sense of trace distance.
Let $\kappa$ be a state such that (i) $\kappa$ is $\ep'$ close to a pure state (i.e., there exists a pure state $\psi$ such that $\|\kappa-\psi\|_1\leq \ep'$), (ii) $\kappa$ is convertible to $\rho'$ by a covariant operation, (iii) $\calC(\kappa)=\calC(\rho')$.
The existence of such $\kappa$ is guaranteed by Lemma~\ref{t:distill-exact} as follows.
First, if $\tilde \kappa$ satisfies (i) and (iii), then $\kappa = \tilde \kappa^{\otimes r}$ with sufficiently large $r$ also satisfies (i) with $\ep'\to r\ep'$ and (iii), as well as (ii) due to Lemma~\ref{t:distill-exact}.
It therefore suffices to ensure the existence of such $\tilde\kappa$ satisfying (i) and (iii) with replacing $\ep'$ by $\ep'/r$.
Such a state for an arbitrary $\ep'>0$ can be constructed by, e.g., purifying $\rho'$ using the auxiliary system with a trivial Hamiltonian and applying the depolarizing channel with sufficiently small noise strength.

We replace $\rho'$ in the above protocol by $\kappa$, and let $\Lambda$ be the covariant operation converting $\rho^{\otimes n}$ to $\kappa^{\otimes m}$.
Due to the condition (i), the correlation between state $\kappa$ and other copies is bounded as $\| \Lambda(\rho^{\otimes n})-\kappa \otimes \Tr_{i}[\Lambda(\rho^{\otimes n})]\|_1 <2\ep' + \sqrt{\ep'/2}$~\cite{Ganardi2023catalytic} for all $i$. 
Take small enough $\ep'$ such that $2\ep'+\sqrt{\ep'/2}<\ep$ and let $\kappa$ be a state satisfying (i)--(iii) for this $\ep'$.
Let $\calE$ be a covariant operation $\calE(\kappa)=\rho'$ ensured by the condition (ii). 
Then, $\tau:=\calE^{\otimes m}\circ\Lambda(\rho^{\otimes n})$ is the desired final state because 
\bal
 \Tr_{\backslash i}\tau =\calE(\Tr_{\backslash i}\Lambda(\rho^{\otimes n})) = \calE(\kappa) = \rho' 
\eal
for all $i$, and 
\bal
 \|\tau - \rho'\otimes \Tr_i\tau\|_1&=\|\calE^{\otimes m}\circ\Lambda(\rho^{\otimes n}) - \calE^{\otimes m}(\kappa\otimes \Tr_i\Lambda(\rho^{\otimes n})\|_1\\
 &\leq \|\Lambda(\rho^{\otimes n}) - \kappa\otimes \Tr_i\Lambda(\rho^{\otimes n})\|_1\\
 &< \ep
\eal
for all $i$, where we used the contractivity of the trace norm under $\calE^{\otimes m}$.
\end{proof}

We remark that our protocol requires exact catalysts, not approximated ones, unlike the previous result on entanglement~\cite{Ganardi2023catalytic}, where the catalyst can be an approximated one.
This difference comes from the difference between a correlated single catalyst and multiple marginal catalysts.
If we employ a single correlated catalyst with some error, this error does not increase through the above process.
On the other hand, if we employ multiple marginal catalysts with some errors, these errors may increase through the above process.
To avoid this trouble, we should prepare catalysts without any small errors.


\subsection{Consistency with zero coherence distillation rate} \label{subsec:consistency with zero rate}

Theorem~\ref{thm:asymptotic go} ensures that one can produce a state with an arbitrary size whose marginal is arbitrarily close to a pure state---which includes the maximally coherent state $\ket{+}=\frac{1}{\sqrt{2}}(\ket{0}+\ket{1})$---with an arbitrarily small correlation between one and the other subsystems.
One may wonder how this would be consistent with the vanishing asymptotic distillation rate $R(\rho\to \phi)=0$ for every full-rank state $\rho$ and pure coherent state $\phi$~\cite{Mar20}, as it might appear that the final state with arbitrarily small correlation should also be arbitrarily close to the tensor product of $\phi$, leading to the contradiction with the zero asymptotic distillation rate. 
Here, we show that these two results are consistent. 

The key observation is that even if a state $\tau_m$ satisfies ${S'}^{\otimes m}$ with $\|\Tr_{\backslash i} \tau_m - \phi\|_1 < \ep$ for all $i$ and $\|\tau_m-\phi\otimes\Tr_i \tau\|_1<\ep$ with small $\ep>0$, $\tau_m$ can generally be far from $\phi^{\otimes m}$ for large $m$.
Here we bound both the error in each subsystem and the amount of correlation by the same variable $\ep$ for notational simplicity.
We denote the distance between $\tau_m$ and $\phi^{\otimes m}$ by
\bal
 \|\tau_m-\phi^{\otimes m}\|_1 := f(m,\ep).
\eal
Then, in order for $\{\tau_m\}_m$ to be a valid family of states for the (standard) asymptotic transformation by taking small $\ep$, it is required that
\bal
 \lim_{\ep\to 0}\lim_{m\to\infty} f(m,\varepsilon) = 0.
\eal
However, this requirement does not hold in general. For instance, let $\phi=\dm{+}$ and consider a class of states defined by 
\bal
\tau_{m} = (1-\delta)\left[(1-\ep/2)\dm{+}+ \frac{\ep}{2} \dm{-}\right]^{\otimes m} + \frac{\delta}{2}(\dm{+}^{\otimes m} + \dm{-}^{\otimes m}).
\label{eq:example final state}
\eal
This satisfies $\|\Tr_{\backslash i}\tau_m-\dm{+}\|_1\leq \ep$ and $\|\tau_m - \dm{+}\otimes \Tr_i \tau\|_1\leq \ep$ for every $i$ for sufficiently small $\delta$.
On the other hand, 
\bal
 f(m,\ep) = 2[1-(1-\delta)(1-\ep/2)^m - \delta/2].
\eal
which satisfies $\lim_{\ep\to 0}\lim_{m\to\infty}f(m,\ep)=2(1-\delta/2)>0$.

This example shows that the combination of ``good local states'' and ``small correlation'' does not necessarily result in a ``good global state'' and ensures that Theorem~\ref{thm:asymptotic go} is not in contradiction with the zero asymptotic distillation rate. 
We remark that the above example may not be a general form that can be obtained by our protocol. 
Nevertheless, this example is already sufficient to argue that there is no definite inconsistency between our result and the zero asymptotic distillation rate. 


\subsection{Consistency with asymmetry measure}

Besides the zero asymptotic distillation rate, one might also find it strange that an arbitrary number of highly coherent states can be prepared in every subsystem from the viewpoint of resource measures.
Theorem~\ref{thm:asymptotic go} shows that from a state with an arbitrarily small amount of coherence, one could prepare a state whose local state is arbitrarily close to a highly coherent state with an arbitrarily small correlation. 
If we naively think that the small correlation would ensure that the total coherence is approximately the sum of local asymmetry, this appears to lead to the contradiction.

To formalize this concern, consider a tensor-product additive asymmetry measure, i.e., an asymmetry measure $R$ satisfying $R(\otimes_i\,\rho_i) = \sum_i R(\rho_i)$. 
The standard tensor-product additive asymmetry measure includes quantum Fisher information~\cite{Zhang2017detecting} and Wigner-Yanase skew information~\cite{Marvian2014extending,Tak19}, which are in a family of additive asymmetry measures known as metric-adjusted skew informations~\cite{Hansen2008metric,Tak19}.

As in the previous subsection, let $\tau_m=\Lambda(\rho^{\otimes n})$ be a state in ${S'}^{\otimes m}$ such that $\|\Tr_{\backslash i} \tau_m - \phi\|_1 < \ep$ for all $i$ and $\|\tau_m-\phi\otimes\Tr_i \tau\|_1<\ep$ for some pure coherent state $\phi$.
Using the tensor-product additivity of $R$ and the monotonicity of $R$ under covariant operations, we get 
\bal
 R(\rho) = \frac{1}{n}R(\rho^{\otimes n}) \geq \frac{1}{n}R(\Lambda(\rho^{\otimes n})) = \frac{1}{n}R(\tau_m).
 \label{eq:asymmetry measure bound}
\eal
If $R$ is also continuous with respect to the trace distance, Theorem~\ref{thm:asymptotic go} claims that even for a state $\rho$ for which $R(\rho)$ is arbitrarily close to 0, \eqref{eq:asymmetry measure bound} holds for an arbitrarily large $m$ and arbitrarily small $\ep$ for a sufficiently large $n$. 
This appears a counterintuitive claim, and indeed, if $\tau_m$ had no correlation at all, this would immediately meet the contradiction.
This is because if $\tau_m = \otimes_i \rho'_i$ for some coherent states $\rho'_i$ close to $\phi$, then tensor-product additivity of $R$ would imply $\frac{1}{n}R(\tau_m) = \frac{1}{n}\sum_{i=1}^m R(\rho'_i)\sim \frac{m}{n}R(\phi)$, which could be made arbitrarily large by taking large $m$. 

The above concern is based on the naive observation that (1) $\tau_m$ should be close to $\phi^{\otimes n}$, and (2) $R(\phi^{\otimes m})= m R(\phi)$, therefore (3) $R(\tau_m)$ should be close to $m R(\phi)$, which might cause an inconsistency. 
To see where this argument breaks down, let us consider the continuity of $R$.  
Let $\rho_1$ and $\rho_2$ be two $d$-dimensional states and $t:=\|\rho_1-\rho_2\|_1$.
The continuity can be formalized as  
\bal
 |R(\rho_1) - R(\rho_2)|\leq g(d,t) 
\eal
where $g$ is a continuous bounded function for $t\in[0,1]$ such that $\lim_{t\to 0}g(d,t)=0$.
Let us take $m=rn$ and $t_{r,n,\ep}:=\|\tau_{rn}-\phi^{\otimes rn}\|_1$. 
Then, the tensor-product additivity and the continuity of $R$ gives
\bal
 R(\rho) &\geq \frac{1}{n} R(\tau_m)\\
 &\geq rR(\phi)- \frac{g(2^{rn},t_{r,n,\ep})}{n}.
 \label{eq:consistency inequality}
\eal
This inequality prohibits the case of $\lim_{\ep\to 0}\lim_{n\to\infty}\frac{g(2^{rn},t_{r,n,\ep})}{n}=0$, because by taking sufficiently small $\ep$ and sufficiently large $n$ (that depends on the chosen $\ep$), the right-hand side of \eqref{eq:consistency inequality} could become arbitrarily close to $rR(\phi)$, which would violate \eqref{eq:consistency inequality} by taking large enough $r$.

To avoid $\lim_{n\to\infty}\frac{g(2^{rn},t_{r,n,\ep})}{n}=0$, the function $g$ must scale with $d$ at least as $\log d$ (i.e., linear in $n$).
If $g$ grows faster than $\log d$, then we get $\lim_{n\to\infty} g(2^{rn}, t_{r,n,\ep})/n=\infty$, making \eqref{eq:consistency inequality} consistent.
Therefore, the only case that the consistency with \eqref{eq:consistency inequality} might be in question is when $g$ grows with $\log d$ and gives $g(2^{rn},t_{r,n,\ep})/n =:h(t_{r,n,\ep})$ satisfying $\lim_{x\to 0}h(x)=0$. 
When this holds, the function $R$ is said to be \emph{asymptotically continuous}~\cite{Synak-Radtke2006on_asymptotic}.
In this case, the issue may arise if $\lim_{\ep\to 0}\lim_{n\to\infty}t_{r,n,\ep} = 0$ for an arbitrary $\tau_{rn}$ such that $\|\Tr_{\backslash i}\tau_{rn}-\phi\|_1\leq \ep$ for all $i$ and $\|\tau_{rn} - \phi\otimes \Tr_i \tau_{rn}\|_1\leq \ep$. 
However, this is not the case in general as we show in the previous section with the example in \eqref{eq:example final state}.
This confirms that the apparent inconsistency with additive asymmetry measures actually does not arise. 

In fact, the above argument can be employed to obtain a general continuity property of weakly tensor-product additive functions, which may be of independent interest. 
Let $f$ be a map from quantum states to real numbers satisfying $|f(\rho_1)-f(\rho_2)|\leq g(d,t)$ for arbitrary $d$-dimensional quantum states $\rho_1$ and $\rho_2$, where $t=\|\rho_1-\rho_2\|_1$ and $g$ is a continuous bounded function for $t\in[0,1]$ such that $\lim_{t\to 0}g(d,t)=0$. 
We say that $f$ is \emph{more than asymptotically continuous}~\cite{Coladangelo2019additive} if $\lim_{d\to\infty}\frac{g(d,t_d)}{\log d}=0$ for an arbitrary sequence $\{t_d\}_d$ such that $t_d\in[0,1]$.

\begin{pro}\label{pro:more than asymptotically continuous}
Let $f$ be an arbitrary non-constant map from quantum states to real numbers that is weakly tensor-product additive, i.e., $f(\rho^{\otimes n})=nf(\rho)$ for an arbitrary state $\rho$ and a positive integer $n$. Then, $f$ cannot be ``more than asymptotically continuous''. 
\end{pro}
\begin{proof}
    Since $f$ is not a constant map, there exist states $\rho$ and $\sigma$ such that $f(\rho)<f(\sigma)$. Let $d$ be the dimension of $\rho$ and $\sigma$, and let $t_n := \|\rho^{\otimes n}-\sigma^{\otimes n}\|_1$. Then, we get 
    \bal
     f(\rho) &= \frac{1}{n}f(\rho^{\otimes n})\\
     &\geq \frac{1}{n}f(\sigma^{\otimes n}) - \frac{g(d^n,t_n)}{n}\\
     &= f(\sigma) - \frac{g(d^n,t_n)}{n}
    \eal
    where the first and the last equalities are due to the weak tensor-product additivity.
    If $f$ is more than asymptotically continuous, we have $\lim_{n\to\infty}g(d^n,t_n)/n = 0$. 
    This then implies $f(\rho)\geq f(\sigma)$ by taking the $n\to\infty$ limit on both sides, which is a contradiction.
\end{proof}
We remark that Ref.~\cite{Coladangelo2019additive} showed---with a different technique---that tensor-product additive, permutationally invariant non-constant functions cannot be more than asymptotically continuous. 
Proposition~\ref{pro:more than asymptotically continuous} extends it to all \emph{weakly} tensor-product additive non-constant functions, which may not necessarily be permutationally invariant.  


\subsection{No-go theorem for asymptotic marginal transformation (Theorem~\ref{thm:asymptotic})} \label{subsec:asymptototic no go appendix}

Theorem~\ref{thm:asymptotic go}
states that if $\calC(\rho')\subseteq \calC(\rho)$ is satisfied, the asymptotic marginal transformation rate $\tilde R(\rho\to \rho')$ becomes unbounded.
Here, we show its opposite: $\calC(\rho')\nsubseteq \calC(\rho)$ implies $\tilde R(\rho\to \rho')=0$, showing that the condition in terms of the set of resonant coherent modes serves as a sharp threshold between the infinite and zero transformation rates.

\begin{thm}[The second part of Theorem~\ref{thm:asymptotic} in the main text]\label{thm:asymptotic no-go}
    If $\mathcal{C}(\rho')\not\subseteq\mathcal{C}(\rho)$, even a single copy of $\rho'$ cannot be prepared from any number of copies of $\rho$ with arbitrarily small error, which particularly implies that the asymptotic marginal transformation rate becomes zero: $\tilde R(\rho\to\rho')=0$. 
\end{thm}

This result can be formalized as follows.

\begin{thm}\label{thm:asymptotic no-go single copy appendix}
Suppose two states $\rho$ and $\rho'$ satisfy $\calC(\rho')\nsubseteq \calC(\rho)$. Then, there exists $\varepsilon>0$ such that for every sequence $\{\Lambda_n\}_n$ of covariant operations,  $\|\Lambda_n(\rho^{\otimes n})-\rho'\|_1 > \varepsilon$ holds for all $n$.
\end{thm}

The main idea behind the proof is that the modes of asymmetry defined in Definition~\ref{def:mode of asymmetry} cannot be created by a covariant operation~\cite{Marvian2014modes}.

\begin{proof}[Proof of Theorem~\ref{thm:asymptotic no-go single copy appendix}]
Suppose, to the contrary, that $\rho$ and $\rho'$ satisfy $\calC(\rho')\not\subseteq\calC(\rho)$ but there exists a series $\{\Lambda_n\}_n$ of covariant operations such that for every $\ep>0$, there is a sufficiently large $n$ satisfying $\|\Lambda_n(\rho^{\otimes n})-\rho'\|_1<\ep$.

Take sufficiently small $\ep>0$ such that $\mathcal{C}(\rho')\subseteq\mathcal{C}\left(\Lambda_n(\rho^{\otimes n})\right)$. 
There always exists such $\ep$ because each entry of the density matrix is continuous with respect to the change in the density matrix with trace distance, and thus every non-zero mode remains as a non-zero mode with a sufficiently small perturbation (while zero modes could turn into non-zero modes under arbitrarily small perturbation.)

This means that, together with the assumption $\calC(\rho')\not\subseteq\calC(\rho)$, there exists $n$ and a covariant operation $\Lambda_n$ such that $\mathcal{C}\left(\Lambda_n(\rho^{\otimes n})\right)\nsubseteq \mathcal{C}(\rho)$.
By Definitions~\ref{def:mode of asymmetry}~and~\ref{def:set of resonant coherent modes} of modes of asymmetry and sets of resonant coherent modes, $\calD(\rho_1)\subseteq\calC(\rho_2)$ implies $\calC(\rho_1)\subseteq\calC(\rho_2)$ for arbitrary states $\rho_1$ and $\rho_2$.
This observation leads to $\mathcal{D}\left(\Lambda_n(\rho^{\otimes n})\right)\nsubseteq \mathcal{C}(\rho)$, because if $\mathcal{D}\left(\Lambda_n(\rho^{\otimes n})\right)\subseteq \mathcal{C}(\rho)$ held contrarily to our claim, then $\mathcal{C}\left(\Lambda_n(\rho^{\otimes n})\right)\subseteq \mathcal{C}(\rho)$ would also hold, contradicting our previous finding $\mathcal{C}\left(\Lambda_n(\rho^{\otimes n})\right)\nsubseteq \mathcal{C}(\rho)$.

Notice that 
\bal
\calD(\rho^{\otimes m})=\lset x \sbar x=\sum_{(i,j)\in \calM} n_{ij}\Di _{ij} : n_{ij}\in \bbZ, \ \sum_{(i,j)}\abs{n_{ij}}\leq m \rset \subseteq \calC(\rho)
\eal
for any $m$.
This, together with $\mathcal{D}(\Lambda_n(\rho^{\otimes n}))\nsubseteq\mathcal{C}(\rho)$, implies $\mathcal{D}\left(\Lambda_n(\rho^{\otimes n})\right)\nsubseteq \mathcal{D}(\rho^{\otimes n})$. 
However, this contradicts the fact that the modes of asymmetry cannot be created by a covariant operation~\cite{Marvian2014modes}, i.e., $\calD(\Lambda(\rho))\subseteq \calD(\rho)$ for every state $\rho$ and covariant operation $\Lambda$.
\end{proof}

We also obtain the no-go theorem for asymptotic exact marginal transformation. 
The case when $\calC(\rho')\not\subseteq\calC(\rho)$ can be shown as a direct consequence of Theorem~\ref{thm:asymptotic no-go single copy appendix}. 
We can also add another constraint when $\rho$ is pure. 

\begin{thm}
    If either (1) $\mathcal{C}(\rho')\not\subseteq\mathcal{C}(\rho)$ or (2) $\rho'$ is pure coherent state and $\rho$ is full rank, there is no covariant operation $\Lambda$ such that $\Lambda(\rho^{\otimes n})=\rho'$ for any integer $n$. 
    As a result, we get $\tilde R^0(\rho\to\rho')=0$
\end{thm}
\begin{proof}
    Theorem~\ref{thm:asymptotic no-go single copy appendix} implies the part for the case when $\calC(\rho')\not\subseteq \calC(\rho)$. 
    The case when $\rho'$ is a pure coherent state and $\rho$ is full rank is prohibited by the fact that pure coherent states have the diverging purity of coherence~\cite{Mar20} while full rank states have the finite purity of coherence.
\end{proof}


\section{Catalytic transformation} \label{sec:catalytic}

\subsection{Preliminaries}

Another approach to enhance the desired transformation is to use an auxiliary state that aids the transformation.
In particular, when the auxiliary state is returned to the original form, it is called a \emph{catalyst}. 
The most direct framework for such a transformation is to transform a state $\rho\otimes c$ to the final state $\rho'\otimes c$, where the catalyst $c$ can be reused for another transformation. 
We particularly call this \emph{product catalyst}, as the transformation keeps the product structure in the final state. 
It turned out that the product catalyst is able to enhance the feasible transformation in various settings, such as entanglement transformation~\cite{Jonathan1999entanglement-assisted,Klimesh2007inequalities,Turgut2007catalytic_transformations}, quantum thermodynamics~\cite{Brandao15second_laws}, speakable coherence~\cite{Bu2016catalytic}, and magic state transformation in the context of fault-tolerant quantum computation~\cite{Campbell2011catalysis}. 

If one focuses on the infinite reusability of the catalyst, the above transformation framework can be extended. 
After one round of the transformation, if it is promised that a fresh initial state (uncorrelated with the previous final state) is prepared for the next transformation, what only matters for the next transformation is the reduced state in the catalytic system --- as long as the reduced state in the catalytic system remains intact from the initial form, it can still be reused indefinitely. 
This is called \emph{correlated catalyst}, which we formally define as follows.

\begin{dfn}[Correlated-catalytic transformation]
We say that state $\rho$ in system $S$ is convertible to state $\rho'$ in system $S'$ by a correlated-catalytic transformation if there exists a finite-dimensional catalytic system $C$ with a state $c$ and a free operation $\Lambda\in\bbO: S\otimes C\to S'\otimes C$ such that 
\eq{
\tau=\Lambda(\rho\otimes c), \hspace{10pt} \Tr_C[\tau]=\rho', \hspace{10pt} \Tr_{S'}[\tau]=c.
}
\end{dfn}
The correlated-catalytic transformation has been proven to be effective in several physical settings.  
A prominent example is quantum thermodynamics, where a correlated catalyst enables us to recover the second law of thermodynamics in a conventional form~\cite{Muller2018correlating, Shiraishi2021quantum}.
It has also been discovered that similar observations can be made to the entanglement transformation~\cite{Lipka-Bartosik2021catalytic,Kondra2021catalytic} and a general class of resource theories~\cite{TS22}.

However, when it comes to the covariant operations, several previous works indicated that the use of a catalyst may not help much for the coherence transformation. 
First, it is known that a pure product catalyst does not enhance the transformation at all in the covariant operations~\cite{Marvian2013theory_manipulations,Ding2021amplifying_asymmetry}. 
In addition, as for the correlated catalyst, it was shown that one could not create any coherence from an incoherent state with a correlated-catalytic covariant operation, i.e., if $\rho$ is incoherent, then so is $\rho'$~\cite{Lostaglio2019coherence_asymmetry,Marvian2019no-broadcasting} (see Theorem.~\ref{t:no-broadcast} for its precise statement).

\subsection{Arbitrary coherence transformation with correlated catalysts (Theorem~\ref{thm:correlated catalyst achievability})}

Contrarily to the aforementioned implications, we here show that correlated catalysts can provide a dramatic operational power, and the limitations imposed on the correlated-catalytic covariant transformation shown in Refs.~\cite{Lostaglio2019coherence_asymmetry,Marvian2019no-broadcasting} (coherence no-broadcasting theorem) is exceptional for the incoherent input states. 

There exists a standard technique to reduce asymptotic transformations to correlated-catalytic transformations.
This technique for non-exact asymptotic transformations was first discussed by Shiraishi and Sagawa~\cite{Shiraishi2021quantum}.
This paper has commented its general applicability, and a number of papers have applied this technique to various resource theories including entropy conjecture~\cite{Wilming2021entropy}, entanglement~\cite{Kondra2021catalytic, Ganardi2023catalytic}, and teleportation~\cite{Lipka-Bartosik2021catalytic}.
The statement in a general form is presented by the authors (Proposition 4 in Ref.~\cite{TS22}).

For our purpose, we here express a slightly generalized version of the statement. 
The proof is the same as that in Ref.~\cite{TS22} and was employed in Ref.~\cite{Wilming2022correlations, Ganardi2023catalytic}.

\begin{pro}[Slight generalization of Ref.~\cite{TS22}]\label{t:asymptotic-catalytic}
Consider a resource theory such that a set of free operations includes the relabeling of a classical register and the conditioning of free operations by a classical register.
Then, if there is a free transformation $\Lambda: S^{\otimes n}\to S^{\otimes n}$ which maps $\rho^{\otimes n}$ to $\tau$ with $\sigma:=\frac1n \sum_{i=1}^n \Tr_{\bcs i}[\tau]$, then there exists a correlated-catalytic free transformation mapping $\rho$ to $\sigma$.
\end{pro}

We here assume, without loss of generality, that the initial state $\rho$ and the final state $\sigma$ are states in the same system $S$. 
This is because if the system $S'$ for the final state was different from the input system $S$, one could consider an enlarged system $S\oplus S'$ as its input and output system and simply call it $S$. 
This modification is harmless since an embedding process is a covariant operation and we can make possible errors in the final state inside the $S'$ part.

For completeness, we here present the construction of the desired correlated-catalytic transformations.
Suppose that a covariant operation $\calE$: $S^{\otimes n}\to S^{\otimes n}$ converts $\rho^{\otimes n}$ to $\tau$ with $\sigma:=\frac1n \sum_{i=1}^n \Tr_{\bcs i}[\tau]$.
Then, defining the reduced state of copies from the first copy to the $i$-th copy as 
\eq{
\tau_i:=\Tr_{C_{i+1},\ldots , C_n}[\tau]\in S^{\otimes i},
}
we construct the catalytic system $C=S^{\otimes (n-1)}\otimes R$ with state $c$ as
\eq{
c:=\frac1n \sum_{k=1}^n \rho^{\otimes (k-1)}\otimes \tau_{n-k}\otimes \ket{k}\bra{k}_R.
}
Here $R$ is a classical register system spanned by $\{\ket{k}\}_{k=1}^n$.
The initial state of the composite system is written as
\eq{
\rho\otimes c=\frac1n \sum_{k=1}^n \rho^{\otimes k}\otimes \tau_{n-k}\otimes \ket{k}\bra{k}_R.
}
For later discussion, we say that the first copy corresponds to system $S$, and the latter $k-1$ copies correspond to catalyst $C$.

Our free operation consists of two steps.
In the first step, we apply $\Lambda$ if the classical register is $\ket{n}\bra{n}$ and leave the system otherwise.
The resulting state is
\eq{
\frac1n \( \sum_{k=1}^{n-1} \rho^{\otimes k}\otimes \tau_{n-k}\otimes \ket{k}\bra{k}_R +\tau\otimes \ket{n}\bra{n}_R \) .
}
In the second step, we relabel the classical register as $k\to k+1$ for $k\neq n$ and $n\to 1$, which results in
\eq{
\frac1n \sum_{k=1}^{n} \rho^{\otimes (k-1)}\otimes \tau_{n-k+1}\otimes \ket{k}\bra{k}_R .
}
Finally, by regarding the first $n-1$ copies as catalyst $C$ and the last copy as system $S$, the state of the catalyst returns to its own state:
\bal
\Tr_n \[ \frac1n \sum_{k=1}^{n} \rho^{\otimes (k-1)}\otimes \tau_{n-k+1}\otimes \ket{k}\bra{k}_R\] &=\frac1n \sum_{k=1}^{n} \rho^{\otimes (k-1)}\otimes \Tr_n[\tau_{n-k+1}] \otimes \ket{k}\bra{k}_R\\
&=\frac1n \sum_{k=1}^{n} \rho^{\otimes (k-1)}\otimes \tau_{n-k}\otimes \ket{k}\bra{k}_R\\
&=c,
\eal
and the state of the system is $\sigma$:
\eq{
\Tr_{\bcs n, R} \[ \frac1n \sum_{k=1}^{n} \rho^{\otimes (k-1)}\otimes \tau_{n-k+1}\otimes \ket{k}\bra{k}_R\] =\frac1n \sum_{k=1}^{n} \Tr_{\bcs n}[\tau_{n-k+1}]=\frac1n \sum_{k=1}^n \Tr_{\bcs k}[\tau]=\sigma.
}

\bigskip

Now we state our second main result on correlated-catalytic transformations, stating that any coherent state is convertible to any full-rank state with a correlated catalyst.

\begin{thm}[Theorem~\ref{thm:correlated catalyst achievability} in the main text]\label{thm:correlated achievability approx and exact}
Let $\rho$ and $\rho'$ be arbitrary states such that $\calC(\rho')\subseteq \calC(\rho)$.
Then, for any accuracy $\delta>0$ and correlation strength $\ep>0$, there exists a finite-dimensional catalyst $c$ and a covariant operation $\Lambda$ such that
\eq{
\Lambda(\rho\otimes c)=\tau, \hspace{10pt} \|\Tr_C[\tau]-\rho'\|_1<\delta, \hspace{10pt} \Tr_S[\tau]=c, \hspace{10pt} \|\tau-\rho'\otimes c\|_1<\ep.
}
Moreover, if $\rho'$ is full rank, the transformation can be made exact, i.e., $\Tr_C[\tau]=\rho'$.
\end{thm}

\begin{proof}
If we do not require the smallness of correlation, this is a direct consequence of Theorem~\ref{thm:asymptotic exact go} and Proposition~\ref{t:asymptotic-catalytic}.
Since one can let the system for the classical register come with the trivial Hamiltonian, in which case relabeling of the classical register is a covariant operation, Proposition~\ref{t:asymptotic-catalytic} can be applied to the case of covariant operations. 
Theorem~\ref{thm:asymptotic exact go} then implies that the exact catalytic transformation is possible for a full-rank target state $\rho'$. 
Since the set of full-rank states is dense in the state space, for every non-full rank state $\rho'$ and every $\delta>0$, there exists a full-rank state that is $\delta$-close to $\rho'$, which shows the statement in the theorem. 

To make the correlation arbitrarily small, we employ the trick that has already been used in the proof of Theorem~\ref{thm:asymptotic exact go}.
Let $\kappa$ be a state such that (i) $\kappa$ is $\ep'$ close to a pure state (i.e., there exists a pure state $\psi$ such that $\|\kappa-\psi\|_1\leq \ep'$), (ii) $\kappa$ is convertible to $\rho'$ by a covariant operation, (iii) $\calC(\kappa)=\calC(\rho')$.
Applying Proposition~\ref{t:asymptotic-catalytic} to the marginal-asymptotic transformation from $\rho$ to $\kappa$ with transformation rate $R=1$ ensured by Theorem~\ref{thm:asymptotic exact go}, we obtain a correlated-catalytic transformation from $\rho$ to $\kappa$.
By taking sufficiently small $\ep'$ in the condition (i), the correlation between state $\kappa$ and the catalyst is bounded from above by $\ep$, which means that $\tau'=\Lambda(\rho\otimes c)$ satisfies $\|\tau'-\kappa\otimes c\|_1<\ep$.
Finally, applying a covariant operation that transforms $\kappa$ to $\rho'$, we arrive at the desired correlated-catalytic transformation with the correlation less than $\ep$.
Here, we used the fact that an operation acting on only the system (not on the catalyst) does not increase the correlation between the system and the catalyst.
\end{proof}

The anomalous power of correlated-catalytic transformations for a two-level system was first argued by Ref.~\cite{Ding2021amplifying_asymmetry}, while as pointed by Ref.~\cite{TS22}, there was a gap in their proof.
The same paper~\cite{TS22} conjectured that the condition $\calC(\rho')\subseteq \calC(\rho)$ would be the necessary and sufficient condition for arbitrary amplification of coherence with a correlated catalyst. 
Theorem~\ref{thm:correlated achievability approx and exact} solves the sufficient part of this conjecture, together with the unproved claim in Ref.~\cite{Ding2021amplifying_asymmetry}, in the affirmative. 
In the next section, we discuss the necessary part of this conjecture and present the significant step toward the full resolution of this problem. 


\subsection{Mode no-broadcasting (Theorem~\ref{thm:correlated catalyst converse})}

We here investigate the limitation of correlated-catalytic transformations with covariant operations.
We first describe an important result known as the coherence no-broadcasting theorem found by Lostaglio and M\"{u}ller~\cite{Lostaglio2019coherence_asymmetry} and Marvian and Spekkens~\cite{Marvian2019no-broadcasting} independently.

\bthm{[Coherence no-broadcasting~\cite{Lostaglio2019coherence_asymmetry, Marvian2019no-broadcasting}]\lb{t:no-broadcast}
Let $\rho$ be an incoherent state on system $S$.
Consider a correlated-catalytic transformation from $\rho$ to $\rho'$ by a covariant operation $\Lambda$ on $SC$ with catalyst $C$: $\Tr_S[\Lambda(\rho\otimes c)]=c$.
Then, the final state of the system $\rho'=\Tr_C[\Lambda(\rho\otimes c)]$ is still an incoherent state.
}

This theorem applies only when an initial state of the system is completely incoherent.
However, it is highly plausible that even if an initial state has some coherence on some modes, this coherence provides no advantage to create coherence on a mode that is only irrationally related to theirs.
This intuition is indeed true, and we can prove the following theorem, which is an extension of the above coherence no-broadcasting theorem to the level of each mode.
To state our finding, we slightly extend the definition of a set of resonant coherent modes. 

For a given set $\calS$ of real numbers, we say that a real number $x$ can be written by a \emph{rational-linear combination} of the elements in $\calS$ if there exists a set $\{a_j\}_j$ of rational numbers such that $x=\sum_j a_j s_j$ for elements $s_j\in\calS$. 
We then define $\calC'(\rho)$ as the set of all real numbers that can be written by a rational linear combination of the set $\calD(\rho)$ of modes of asymmetry defined in Definition~\ref{def:mode of asymmetry}.

\begin{dfn}[Set of rational coherent modes]
We define a set of rational coherence modes denoted by $\calC'(\rho)$ as rational-linear combinations of non-zero coherent modes of $\rho$:
\eq{
\calC'(\rho):=\lset x \sbar x=\sum_{\Di\in \calD(\rho)} a_{\Di}\Di  : a_{\Di}\in \bbQ\rset.
}
\end{dfn}

We then show the following no-go theorem. 

\bthm{[Theorem~\ref{thm:correlated catalyst converse} in the main text: Mode no-broadcasting (weak version)]\label{t:no-go-catalytic}
Consider a correlated-catalytic transformation from $\rho$ to $\rho'$ through a covariant operation $\Lambda$ on $SC$ with catalyst $C$: $\Tr_S[\Lambda(\rho\otimes c)]=c$.
Then, the final state of the system $\rho'=\Tr_C[\Lambda(\rho\otimes c)]$ has no coherence on a mode that is only irrationally related to coherent modes of $\rho$, i.e., $\calC'(\rho')\subseteq \calC'(\rho)$.
}

Namely, if $\calC'(\rho')\nsubseteq \calC'(\rho)$, then no correlated-catalytic transformation maps $\rho$ to $\rho'$.

To show this, we first remark that the input system $S$ and the output system $S'$ of a covariant operation can generally be different. However, as pointed out in Ref.~\cite{Mar-thesis}, one can model such a transformation with another covariant operation with the same input and output systems by considering an extended space $\calH_{\rm in}\oplus\calH_{\rm out}$, where $\calH_{\rm in, out}$ are the Hilbert spaces underlying the systems $S$ and $S'$ respectively.
Therefore, we consider $S$ as the system on the space that is already extended in this way and focus on a covariant operation such that $SC$ is its input and output space.

Our proof of Theorem~\ref{t:no-go-catalytic} is based on the proof by contradiction.
We suppose contrarily that there exists a correlated-catalytic transformation from $\rho$ to $\rho'$ despite $\calC'(\rho')\nsubseteq \calC'(\rho)$.
Our goal is to construct a protocol violating the coherence no-broadcasting theorem.

To deal with any system in a unified framework, we embed the main and the catalytic systems in a product of {\it ladder systems} whose energy levels form an infinite ladder.

\bdf{[Ladder system]
A ladder system with energy interval $\Di$ denoted by $L(\Di)$ has states labeled by two integers $(n,a)$ with $n\in \bbZ$ and $a\in \bbN$.
The state $\ket{n,a}$ is an energy eigenstate with energy $n\Di$.
The label $a$ distinguishes degenerate energy eigenstates.
}

Suppose that the main system $S$ and catalytic system $C$ can be embedded into the collection of the ladder systems.
We consider this extended ladder system as our system $S$ and write it as $S=L_S(\Di_1)\otimes L_S(\Di_2)\otimes \cdots =:L_S(\bsDi)$ with abbreviation $\bsDi=(\Di_1,\Di_2,\ldots)$.
Similarly, we set the catalytic system as $C=L_C(\bsDi)$. 
For later convenience, we write a subsystem whose energy interval is multiples of $\Di_i$ as $X(\Di_i):=L_S(\Di_i)\otimes L_C(\Di_i)$.
We also denote by $X(\bsDi):=L_S(\bsDi)\otimes L_C(\bsDi)$.

We remark that this embedding is always possible when all energies for $S$ and $C$ can be written as integer-linear combinations of $\bsDi$. For instance, suppose that an energy $E$ for (either the main or catalytic) system is written as $E=\sum_j n_j \Di_j$ for some integers $\{n_j\}_j$ and elements $\{\Di_j\}_j$ of $\bsDi$.
Then, the energy eigenstates $\ket{E,\alpha}$ (where $\alpha$ distinguishes the degeneracy) can be mapped as $\ket{E,\alpha}\to\otimes_j \ket{n_j,\alpha}_{\Di_j}$ where $\ket{n_j,\alpha}_{\Di_j}$ is an energy eigenstate in $L(\Di_j)$. 
By construction, $\otimes_j \ket{n_j,\alpha}_{\Di_j}$ is an energy eigenstate of the extended ladder system with energy $\sum_j n_j \Di_j$.

In the following, we say that a set $\calS$ is \emph{rational-linearly independent} if, for all elements $x_i\in\calS$, there is no set $\{a_{j}\}_j$ of rational numbers such that $x_i = \sum_{j\neq i} a_{j} x_j$.
Our key observation is that if a covariant operation on $X(\bsDi)$ exists and all $\bsDi$ is rational-linearly independent, then the value of $\bsDi$ is in fact irrelevant.
To describe this, we denote by $\rho[\bsDi]$ a quantum state on $S=L_S(\bsDi)$ whose density matrix is $\rho$.
Similarly, we denote by $c[\bsDi]$ and $\tau[\bsDi]$ quantum states on $C=L_C(\bsDi)$ and $SC=X(\bsDi)$, respectively.

\begin{lm}\label{t:irrational-tune}
Let $\Di_1, \Di_2, \ldots$ be energy intervals that are rational-linearly independent.
Suppose that a state $\tau[\bsDi]$ on $SC=X(\bsDi)$ is convertible to $\tau'[\bsDi]$ on $SC$ by a covariant operation.
Then, for any $\bsDi'$ state $\tau[\bsDi']$ on $X(\bsDi')$ is also convertible to $\tau'[\bsDi']$ on $X(\bsDi')$ by a covariant operation.
\end{lm}

\begin{proof}
By definition, a covariant operation $\Lambda$ on $SC$ can be expressed by using an auxiliary system $A=L_A(\bsDi)$ with its incoherent state $\eta$ as
\eq{
\Lambda(\tau)=\Tr_{A'}[U(\tau\otimes \eta)U^\dagger],
}
where $U$ is an energy-conserving unitary on $SCA$.
Owing to the energy conservation and rational-linear independence of energy intervals, energy change in $S$ or $C$ occurs only in each $L_S(\Di_i)\otimes L_C(\Di_i)\otimes L_A(\Di_i)$, since the energy change with a multiple of $\Di _i$ cannot be compensated by any rational-linear combination of $\Di_j$'s with $j\neq i$.
In other words, using a set of operators $\{K_i^j\}_j$ acting on $L_S(\Di_i)\otimes L_C(\Di_i)\otimes L_A(\Di_i)$, any energy-conserving unitary $U$ can be expressed in the form of
\eqa{
U=\sum_j \bigotimes_i K_i^j
}{U-decomp}
with
\eq{
[K_i^j, H_{SCA}(\Di_i)]=0
}
for any $i$ and $j$.
Here, $H_{SCA}(\Di_i)$ is a Hamiltonian on $L_S(\Di_i)\otimes L_C(\Di_i)\otimes L_A(\Di_i)$.

As seen from \eref{U-decomp}, $U$ conserves energy regardless of the value of $\Di_i$'s.
This directly implies that the same $U$ for $\bsDi'$ with the initial state of $A$ as $\eta[\bsDi']$ serves as the desired covariant operation.
\end{proof}

We are now in a position to prove Theorem~\ref{t:no-go-catalytic}.

\begin{proof}[Proof of Theorem~\ref{t:no-go-catalytic}]

Let $\rho$ and $\rho'$ be states on the system $S$, and suppose $\calC'(\rho')\not\subseteq\calC'(\rho)$. Then, there exists a mode in $\rho'$ that cannot be written by a rational-linear combination of the modes in $\rho$. 
Let $\Di_0$ be the energy interval for such a mode.
We show that one can construct a set $\bsDi$ of intervals that embeds $SC$ and would lead to the contradiction with coherence no-broadcasting theorem.

To construct the desired $\bsDi$, we start with $\bsDi=\{\Di_0\}$ and add elements to $\bsDi$ step by step.
Let $\overline{\calD}(\rho)$ be a set of modes where $\rho$ does not have coherence, which is a complement of a set $\calD(\rho)$ of coherent modes. 
Let $\Di_i(\rho)$ be the $i$\,th element in $\cal D(\rho)$ (with an arbitrary order).
We run the following procedure. 
\begin{enumerate}
    \item If $\Di_i(\rho)$ cannot be written by a rational-linear combination of the elements already in $\bsDi$, add $\Di_i(\rho)$ to $\bsDi$. On the other hand, if $\Di_i(\rho) = \sum_j \frac{m_j}{n_j} \Di_j$ for $\Di_j\in\bsDi$ and some integers $m_j$ and $n_j$, we redefine all elements $\Di\in\bsDi$ as $\Di\to \Di/\prod_j n_j$, while not adding $\Di_i(\rho)$ to $\bsDi$. We sequentially apply this procedure for $i=1,2,\dots |\calD(\rho)|$.
    \item We then apply the same procedure for all incoherent modes in $\overline{\calD}(\rho)$. Namely, if an incoherent mode cannot be written as a rational-linear combination of the elements already in $\bsDi$, we add the incoherent mode into $\bsDi$. Otherwise, we redefine the elements in $\bsDi$ by dividing them by some integer. 
    \item We apply the same procedure for energy intervals of the catalytic system $C$.
\end{enumerate}

By construction, the resulting set $\bsDi$ satisfies that (1) all energies in $S$ and $C$ can be written as an integer-linear combination of the elements in $\bsDi$, and thus $SC$ can be embedded in $L_S(\bsDi)\otimes L_C(\bsDi)$ (2) $\bsDi$ is rational-linearly independent (3) no combination of $\Di_0$ and other modes in $\bsDi$ results in a coherent mode of $\rho$.
The condition (3) is confirmed as follows:
If a combination of $\Di_0$ and other modes in $\bsDi$ results in a coherent mode of $\rho$, one could write $\Di_0$ as a rational-linear combination of modes in $\calD(\rho)$, which would contradict the assumption that $\Di_0\not\in\calC'(\rho)$. 
A remarkable point of this construction lies in the fact that defining $\tilde \bsDi := \bsDi\setminus\{\Di_0\}$ so that $\bsDi=\{\Di_0,\tilde\bsDi\}$, any state $\rho$ on $L_S(\Di_0)\otimes L_S(\tilde \bsDi)$ which is incoherent on $L_S(\Di_0)$ (i.e., $\Di_0\not\in\calC'(\rho)$) can be expressed as $\rho = \sum_{i,\alpha,\alpha'} p_i\ketbra{i,\alpha}{i,\alpha'}_{\Di_0}\otimes \sigma_i$, since the coefficient of $\ketbra{i,\alpha}{j,\alpha'}_{\Di_0}$ terms with $i\neq j$ should be zero due to absence of coherence on $L_S(\Di_0)$.
Here, $\sigma_i$ are states on $L_S(\tilde\bsDi)$ and $p_i$ are nonnegative coefficients satisfying $\sum_i p_i=1$.

Then, suppose contrarily that a covariant operation $\Lambda$ on $SC$ converts $\Lambda(\rho\otimes c)=\tau$ with $\Tr_S[\tau]=c$ and $\Tr_C[\tau]$ has non-zero coherence for the mode with energy interval $\Di_0$.
This implies that the reduced state of $\Tr_C[\tau]$ on $L_S(\Di_0)$ has non-zero coherence.
We aim to show that this contradicts the coherence no-broadcasting theorem.

To this end, we introduce a method of {\it complete degeneration}.
By setting $\bsDi'=\{\Di_0,\bszero\}$ in Lemma~\ref{t:irrational-tune}, we have a correlated-catalytic covariant transformation from $\rho[\{\Di_0, \bszero\}]$ to $\rho'[\{\Di_0, \bszero\}]$ with a catalyst $c[\{\Di_0, \bszero\}]$.
Remarkably, the form $\rho = \sum_{i,\alpha,\alpha'} p_i\ketbra{i,\alpha}{i,\alpha'}_{\Di_0}\otimes \sigma_i$ and the fact that $L_S(\bszero)$ has a single energy with fully degenerate energy eigenstates---hence no coherence in any state---the initial state of the system $\rho[\{\Di_0, \bszero\}]$ is an incoherent state.
On the other hand, the final state of the system, $\rho'[\{\Di_0, \bszero\}]$ has non-zero coherence in $L_{S}(\Di_0)$.
In summary, an incoherent state $\rho[\{\Di_0, \bszero\}]$ is converted into a coherent state $\rho'[\{\Di_0, \bszero\}]$ by a covariant operation with a correlated catalyst, which contradicts the coherence no-broadcasting theorem.
\end{proof}

Although we do not have a proof at present, we expect that the condition $\calC'(\rho')\nsubseteq \calC'(\rho)$ in our mode no--broadcasting theorem can be lifted to $\calC(\rho')\nsubseteq \calC(\rho)$.

\begin{cjt}[Mode no-broadcasting (strong version)]
Consider a correlated-catalytic transformation from $\rho$ to $\rho'$ by a covariant operation $\Lambda$ with a catalyst $c$: $\Tr_S[\Lambda(\rho\otimes c)]=c$.
Then, the final state of the system $\rho'=\Tr_C[\Lambda(\rho\otimes c)]$ satisfies $\calC(\rho')\subseteq \calC(\rho)$.

\end{cjt}

The nontrivial part of showing the conjecture is to rule out the possibility of creating non-zero coherence from the coherence on another mode that is rationally related. We leave a thorough investigation for future work.


\section{Asymptotic coherence manipulation with correlated catalyst}

The power of correlated catalyst has mainly been considered in the context of enhancing single-shot transformations, i.e., whether a catalyst could perform the transformation from a single copy of $\rho$ to a single copy of $\rho'$ that is not realizable without the help of catalysts. 

An interesting---yet still much unexplored---question is whether correlated catalysts could enhance the \emph{asymptotic} transformation rate by using correlated catalysts alongside the asymptotic transformation. 
This question was recently raised and studied in the context of entanglement distillation~\cite{Lami2023catalysis,Ganardi2023catalytic}. 
Here, let us formally introduce relevant quantities. 

\begin{dfn}[Asymptotic correlated-catalytic transformation rate]
Let $\rho$ and $\rho'$ be states on systems $S$ and $S'$. 
We say that the rate $r$ is achievable in asymptotic correlated-catalytic transformation if there is a series $\{c_n\}_n$ of finite-dimensional states in some systems $\{C_n\}_n$ and a series $\{\Lambda_n\}_{n}$ of free operations with $\Lambda_n:S^{\otimes n}\otimes C_n\to S'^{\otimes \lfloor rn \rfloor}\otimes C_n$ such that for any $\ep>0$ there exists sufficiently large $N$ and for $n>N$
\bal
\|\Tr_{C_n}\Lambda_n(\rho^{\otimes n}\otimes c_n)-{\rho'}^{\otimes \lfloor rn \rfloor}\|_1<\ep,\quad \Tr_{\backslash C_n}\Lambda_n(\rho^{\otimes n}\otimes c_n) = c_n
\eal
is satisfied.
The \emph{asymptotic correlated-catalytic transformation rate} $R_{\rm cc}(\rho\to\rho')$ is the supremum over the achievable rates. 
\end{dfn}

Analogously to the case of other asymptotic transformations, we can also introduce the asymptotic exact rate. 

\begin{dfn}[Asymptotic exact correlated-catalytic transformation rate]
Let $\rho$ and $\rho'$ be states on systems $S$ and $S'$. 
We say that the rate $r$ is achievable in asymptotic exact correlated-catalytic transformation if there is a series $\{c_n\}_n$ of finite-dimensional states in some systems $\{C_n\}_n$ and a series $\{\Lambda_n\}_{n}$ of free operations with $\Lambda_n:S^{\otimes n}\otimes C_n\to S'^{\otimes \lfloor rn \rfloor}\otimes C_n$ such that there exists sufficiently large $N$ and for $n>N$
\bal
\Tr_{C_n}\Lambda_n(\rho^{\otimes n}\otimes c_n)=&{\rho'}^{\otimes \lfloor rn \rfloor} \\
\Tr_{\backslash C_n}\Lambda_n(\rho^{\otimes n}\otimes c_n) =& c_n.
\eal
The \emph{asymptotic exact correlated-catalytic transformation rate} $R_{\rm cc}^0(\rho\to\rho')$ is the supremum over the achievable rates. 
\end{dfn}

In the context of entanglement distillation, it was found that correlated catalysts cannot enable non-zero distillation rate for positive-partial-transpose (PPT) entangled states~\cite{Ganardi2023catalytic} or cannot increase the distillation rates for distillable entangled states~\cite{Lami2023catalysis}. 
Ref.~\cite{Lami2023catalysis} also showed that, in the setting of speakable coherence~\cite{Baumgratz2014quantifying}---related but different framework from that for superposition of energy eigenstates, which we discussed in this article---distillable coherence or coherence cost does not change with the help of correlated catalysts. 
It was then proposed as an open problem whether correlated catalysts could ever improve asymptotic transformation rates in any physical setting. 

Let us now consider our setting of coherence distillation with covariant operations. 
Recall that the standard asymptotic rate $R(\rho\to\phi)$ of coherence distillation by covariant operations is zero for all full-rank state $\rho$ and pure state $\phi$ (Theorem~\ref{t:distill-impossible}). 
This shows that all full-rank states are ``bound coherent'' states analogous to bound entanglement in the resource theory of entanglement, and the corresponding question is whether correlated catalysts could improve this rate, i.e., whether it is possible to obtain $R_{\rm cc}(\rho\to\phi)>0$. 
Our results answer this question in the most drastic way.

\begin{cor}
    Let $\rho$ be a state in $S$ and $\rho'$ be a state in $S'$ such that $\calC(\rho')\subseteq\calC(\rho)$. Then, $R_{\rm cc}(\rho\to\rho')$ diverges. Moreover, if $\rho'$ is full rank, $R_{\rm cc}^0(\rho\to\rho')$ also diverges. In both cases, the correlation between the main and catalytic systems can be made arbitrarily small.
\end{cor}
\begin{proof}
This is a direct consequence of Theorem~\ref{thm:correlated achievability approx and exact} by taking $\rho^{\otimes n}$ as the initial state and ${\rho'}^{\otimes R n}$ as the target state for an arbitrary $R$. 
The condition in Theorem~\ref{thm:correlated achievability approx and exact} is satisfied because $\calC(\sigma^{\otimes m})=\calC(\sigma)$ for every state $\sigma$ and integer $m$.
\end{proof}


\section{Extension to general resource theories}\label{sec:general resource theories}

\subsection{Asymptotic-marginal and correlated-catalytic free transformation}

The arguments to prove Theorems~\ref{thm:asymptotic exact go}~and~\ref{thm:correlated achievability approx and exact} provide a systematic way of constructing asymptotic-marginal and correlated-catalytic transformations from a marginal catalytic transformation (recall its definition in Definition~\ref{def:marginal catalytic}). 
Notably, what we have employed is only the aforementioned general properties, and other specific properties of quantum coherence are not utilized.
Therefore, these results can directly be extended to general resource theories.
We omit the proofs because they are essentially the same as those for Theorems~\ref{thm:asymptotic exact go}~and~\ref{thm:correlated achievability approx and exact}.

\begin{thm}
Consider a resource theory with set $\bbO$ of free operations. 
Let $\rho$ and $\rho'$ be arbitrary states on $S$ and $S'$.
Suppose that $\rho$ can be transformed to $\rho'$ by marginal-catalytic free transformation, i.e., there exist catalytic systems $C_1,\ldots , C_N$ with state $c_1,\ldots , c_N$ and a free operation $\Lambda\in\bbO: S\otimes C_1\otimes \cdots \otimes C_N\to S'\otimes C_1\otimes \cdots \otimes C_N$ such that $\tau=\Lambda(\rho\otimes c_1\otimes \cdots \otimes c_N)$ satisfies $\Tr_{C_1,\ldots , C_N}[\tau]=\rho'$ and $\Tr_{\bcs C_i}[\tau]=c_i$ for any $1\leq i\leq N$.
Suppose also that there exists a free operation $\calE\in\bbO: S^{\otimes m}\to C_1\otimes \cdots \otimes C_N$ such that $\calE(\rho^{\otimes m})=c_1\otimes \cdots \otimes c_N$ for some integer $m$.

Then, for any $\delta>0$, there exist sufficiently large integers $n$ and $m$ with $\frac mn> 1-\delta$ and a free operation $\calK\in\bbO$ on $S^{\otimes n}\to S'^{\otimes m}$ such that 
\eq{
\Tr_{\bcs i}[\calK(\rho^{\otimes n})]=\rho'
}
for any $1\leq i\leq m$.
\end{thm}

\begin{thm}
Consider a resource theory whose free operations include the relabeling of classical registers and free operations conditioned by classical labels.
Let $\rho$ and $\rho'$ be arbitrary states on $S$ and $S'$ such that $\rho$ can be transformed by marginal-catalytic free transformation.
Suppose also that there exists a free operation $\calE: S^{\otimes m}\to C_1\otimes \cdots \otimes C_N$ such that $\calE(\rho^{\otimes m})=c_1\otimes \cdots \otimes c_N$ for some integer $m$.
Then, there exists a finite-dimensional catalytic system $C$, its state $c$, and a free operation $\calK$ on $SC$ such that
\eq{
\calK(\rho\otimes c)=\tau, \hspace{10pt} \Tr_C[\tau]=\rho', \hspace{10pt} \Tr_S[\tau]=c.
}
\end{thm}

\subsection{Restrictions imposed by resource measures} \label{subsec:restriction measure}

Theorems~\ref{thm:asymptotic go}~and~\ref{thm:correlated achievability approx and exact} appear highly anomalous compared to results in other resource theories.
One may wonder why such apparent amplification enabled by correlation is not seen in other resource theories.
To elucidate the specialty of (unspeakable) quantum coherence, we see these phenomena from the viewpoint of resource measures.

To this end, we recall the limitations imposed on correlated and marginal catalytic transformations.

\begin{pro}[Proposition 3 of Ref.~\cite{TS22}]
Let $\frR$ be a resource measure that is tensor-product additive and superadditive.
Then, $\frR(\rho)\geq \frR(\rho')$ holds if $\rho$ is convertible to $\rho'$ by a correlated-catalytic or a marginal-catalytic free transformation.
\end{pro}

Here, a resource measure $\frR$ is tensor-product additive if $\frR(\rho\otimes \sigma)=\frR(\rho)+\frR(\sigma)$, and is superadditive if a state $\tau$ on a composite system $AB$ satisfies $\frR(\tau)\geq \frR(\Tr_A[\tau])+\frR(\Tr_B[\tau])$.
This theorem directly implies that the existence of even a single resource measure satisfying the above conditions and the nontriviality, i.e., there exist two states $\rho$ and $\rho'$ such that $0<\frR(\rho)<\frR(\rho')$, prohibits arbitrary state conversions by a correlated-catalytic free transformation, since conversion $\rho\to \rho'$ with $\frR(\rho)<\frR(\rho')$ is impossible.

The restriction on the asymptotic marginal transformation is obtained in a similar manner.

\begin{pro}
   Let $\frR$ be a resource measure that is tensor-product additive and superadditive.
Then, $\frR(\rho)\geq \tilde R^0(\rho\to\rho')\,\frR(\rho')$ holds.
\end{pro}
\begin{proof}
By definition of the asymptotic exact marginal transformation rate, for every $\delta>0$, there exists a sufficiently large $n$ and a free operation $\Lambda\in\bbO:S\to S'^{\otimes \lfloor(\tilde R^0 (\rho\to\rho')-\delta) n\rfloor}$ such that $\Tr_{\backslash i}\Lambda(\rho^{\otimes n}) = \rho'$ holds for all $i$.
Using such $n$, we get 
\bal
   \frR(\rho) &= \frac{1}{n}\frR(\rho^{\otimes n})\\ 
   &\geq \frac{1}{n}\frR(\Lambda(\rho^{\otimes n}))\\
   &\geq \frac{\left\lfloor n\left(\tilde R^0(\rho\to\rho')-\delta\right) \right\rfloor}{n} \frR(\rho')\\
   &\geq \frac{\left[ n\left(\tilde R^0(\rho\to\rho')-\delta\right) -1 \right]}{n} \frR(\rho')\\
   &=\left[\left(\tilde R^0(\rho\to\rho')-\delta\right) -1/n\right] \frR(\rho')
\eal
where the first line is due to the tensor-product additivity of $\frR$, the second due to the monotonicity, and the third line due to the superadditivity of $\frR$. 
The statement follows by noting that $\delta>0$ and $1/n$ can be made arbitrarily small by taking a sufficiently large $n$. 
\end{proof}

Most resource theories including entanglement~\cite{CW04, AF04}, quantum thermodynamics~\cite{Wilming2017axiomatic}, speakable coherence~\cite{Baumgratz2014quantifying, XLF15} have such a nontrivial measure. (See, e.g., \cite{Hickey2018quantifying,Kuroiwa2020general} for a couple of exceptions.)
In contrast, any nontrivial faithful measure in the resource theory of (unspeakable) coherence is shown not to be superadditive~\cite{Marvian2019no-broadcasting}.
This distinguishes the resource theory of quantum coherence from other resource theories.

\end{document}